\documentclass[twocolumn,aps,prx,floatfix,superscriptaddress]{revtex4-1}

\usepackage{graphicx}
\usepackage{dcolumn}
\usepackage{epsfig}
\usepackage{amssymb}
\usepackage{amsmath}
\usepackage{natbib}
\usepackage{array}
\usepackage{bbold}
\usepackage{color}

\renewcommand{\vec}[1]{\mathbf{#1}}
\renewcommand{\Re}{\mathop{\mathrm{Re}}}
\renewcommand{\Im}{\mathop{\mathrm{Im}}}

\newcommand{\rot}{\mathop{\mathrm{rot}}}

\newcommand{\vep}{\varepsilon}

\begin{document}

\title{Spin-flip processes and radiative decay of dark intravalley excitons \\
in transition metal dichalcogenide monolayers}

\author{A. O. Slobodeniuk}
\affiliation{Laboratoire National des Champs Magn\'etiques Intenses, CNRS-UGA-UPS-INSA-EMFL, 25 avenue des Martyrs, 38042 Grenoble, France}

\author{D. M. Basko}
\address{Laboratoire de Physique et Mod\'elisation des Milieux Condens\'es, Universit\`e de Grenoble-Alpes and CNRS,
25 rue des Martyrs, 38042 Grenoble, France}

\begin{abstract}
We perform a theoretical study of radiative decay of dark
intravalley excitons in transition metal dichalcogenide
monolayers.
This decay necessarily involves an electronic spin flip.
The intrinsic decay mechanism due to interband spin-flip dipole
moment perpendicular to the monolayer plane, gives a rate about
100--1000 times smaller than that of bright excitons.
However, we find that this mechanism also introduces an energy
splitting due to a local field effect, and the whole oscillator
strength is contained in the higher-energy component, while the
lowest-energy state remains dark and needs an extrinsic
spin-flip mechanism for the decay.
Rashba effect due to a perpendicular electric field or a
dielectric substrate, gives a negligible radiative decay rate
(about $10^7$ times slower than that of bright excitons).
Spin flip due to Zeeman effect in a sufficiently strong
in-plane magnetic field can give a decay rate comparable to
that due to the intrinsic interband spin-flip dipole.
\end{abstract}


\maketitle

\section{Introduction}
\label{Sec:1}

Transition metal dichalcogenides (TMDCs) are layered materials
with the chemical composition MX$_2$, where M~is a transition
metal element (such as molybdenum or tungsten), and X~is a
chalcogen (sulfur, selenium, or tellurium).
The interest to semiconducting TMDCs has been sparked by the
recent discovery of the monolayer MoS$_2$ being a direct-gap
semiconductor, in contrast to its bulk indirect-gap counterpart
\cite{Splendiani2010,Mak2010}.
Atomically thin TMDC monolayers can be extracted from bulk
crystals by exfoliation, similarly to graphene
\cite{Splendiani2010,Mak2010,Novoselov2005},
or grown by molecular beam epitaxy or chemical vapor deposition
\cite{Zhan2012,Liu2012,Zhang2013}.
The optical gap in the visible light range
\cite{Splendiani2010,Mak2010} and tightly bound excitons
\cite{Ramasubramaniam2012,Berkelbach2013,Qiu2013,
Chernikov2014,Ye2014,He2014,Zhu2015},
make them quite promising for optical applications
\cite{Eda2014}.

\begin{figure}
\includegraphics[width=7.4cm]{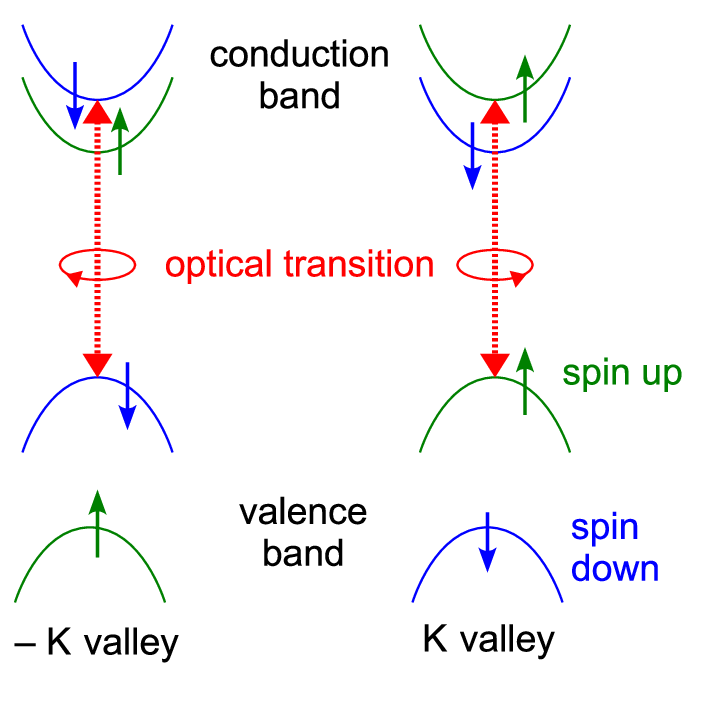}
\caption{\label{fig:TMDCbands} 
A schematic view of the conduction and valence band extrema
of a tungsten-base TMDC monolayer. The solid arrows show the
spin projections. In molybdenum-based compounds, the spin
projections in the conduction band are the opposite.
The dashed arrows represent circularly polarized optical
transitions for $A$ excitons in each valley.
}
\end{figure}

A unique feature of TMDCs is the so-called spin-valley locking
\cite{Xiao2012}.
The conduction and valence band extrema are located at the
two inequivalent $\pm\vec{K}$ points (valleys) of the hexagonal
first Brillouin zone. Absence of inversion symmetry and strong
spin-orbit interaction, originating from $d$-orbitals of the
metal atoms, leads to spin splitting of the bands.
The sign of the splitting is opposite in the two valleys, as
required by the time-reversal symmetry, so the lowest-energy
electron and hole states have opposite spin projections in the
opposite valleys, as shown schematically in Fig.~\ref{fig:TMDCbands}.
It is important that the two valleys can be addressed separately
by optical means. Namely, in each valley optical transitions
with only one of the two in-plane circular polarizations are
allowed~\cite{Xiao2012,Yao2008,Cao2012}.
This opens an exciting perspective of manipulating the
spin/valley degree of freedom optically
\cite{Zeng2012,Mak2012,Sallen2012,Jones2013,Wang2014,Rivera2016,Hao2016}.

The large spin splitting in the valence band (a fraction of eV)
gives rise to two prominent features in the absorption spectrum,
called $A$ and $B$ excitons, formed by a hole in the upper
or lower spin-split valence band, respectively, and an
electron in the conduction band.
The spin splitting in the conduction band is much weaker (on
the meV scale); moreover, its sign depends on the material:
in molybdenum-based compounds, the top valence band and the
lowest conduction band have the same spin projection in the
same valley, while in tungsten-based ones, they are opposite
(Fig.~\ref{fig:TMDCbands})
\cite{Kormanyos2013,Liu2013,Kosmider2013}.
This implies that the lowest-energy $A$ excitons in WX$_2$
involve electronic states either from the same valley but
with opposite spins, or with the same spin but from different
valleys.
As the photons, polarized in the plane, couple only to the
orbital degrees of freedom and carry a very small momentum,
they are decoupled from such excitons, which therefore have
been named ``dark''. This picture provided a natural
explanation of the experimentally observed rise of the
luminescence intensity in WSe$_2$ with increasing temperature
in terms of the increasing thermal population of the
higher-energy bright $A$~excitons
\cite{Arora2015,Withers2015,Wang2015,Zhang2015}.
At sufficiently low temperatures, several peaks at energies
below the neutral and charged bright $A$ excitons have been
seen in the WSe$_2$ luminescence spectrum
\cite{Jones2013,Wang2014,Arora2015,You2015,Mitioglu2015,%
Koperski2015,Srivastava2015,Smolenski2015}.

These observations pose an important question about existence
of possible mechanisms for radiative decay of the dark excitons.
While the radiative decay of bright excitons in TMDCs fits in
the standard picture for 2D excitons
\cite{Agranovich1966,Andreani1990,Andreani1991} and has been
described in several theoretical works
\cite{Glazov2014,Gartstein2015,Palummo2015,Wang2016}, the dark
exciton radiative decay requires unlocking the spin and the
valley degrees of freedom.
Namely, for intervalley dark excitons, a large momentum should
be provided to the electron or to the hole without flipping
its spin, which requires a third body (a phonon, an impurity,
another electron or a hole).
For intravalley dark exciton decay, a spin flip is required.
Investigation of this latter possibility is the subject of
the present work.

In this paper, we give an estimate of the intravalley dark
exciton radiative decay rate due to several spin-flip
mechanisms (see Sec.~\ref{ssec:integrated} for the final
expressions and estimates). One such mechanism could be the
Bychkov-Rashba effect~\cite{Bychkov1984}, as mentioned in
\cite{Dery2015}. It requires breaking of the crystal
symmetry with respect to reflection in the monolayer plane.
This can be achieved by applying an external electric
field, perpendicular to the plane; the corresponding Rashba
coefficient has been calculated in \cite{Kormanyos2014}.
The reflection symmetry can also be broken if the dielectric
environments on the two sides of the monolayer are different.
Having not found any information on the magnitude of this
effect in the literature, we make our own crude estimate;
for a TMDC monolayer on a glass-like substrate with vacuum
above, we obtain an equivalent electric field to be
roughly of the order of 0.1~V/{\AA}.
However, even for this relatively
high field, the resulting radiative rate is about $10^{-7}$
of the bright exciton decay rate~$\Gamma_0$, which makes the
Rashba mechanism totally negligible with respect to other
relaxation processes. The main reason for {this} is
that Rashba spin-flip amplitude is
proportional to the electron momentum, and radiative decay is
possible only in a narrow radiative region of small momenta.
Another way to flip the electron spin is to apply an in-plane
magnetic field and use the Zeeman effect.
It is momentum-independent, and for a quite high but still
realistic field of 30~T we obtain a radiative rate exceeding
$10^{-3}\Gamma_0$.

Finally, we analyzed the intrinsic radiative decay mechanism
due to the interband spin-flip dipole moment perpendicular
to the monolayer plane, mentioned
in \cite{Glazov2014,Echeverry2016}. In fact, it is
the presence of this dipole moment that gives rise to
Bychkov-Rashba coupling in a perpendicular electric
field~\cite{Ochoa2013},
so its magnitude can be deduced from the estimates of
\cite{Kormanyos2014} for the Rashba coefficient.
This enabled us to estimate the associated radiative decay
rate as $\sim(10^{-2}-10^{-3})\Gamma_0$.
However, we also found that the same out-of-plane
interband dipole gives rise to a Coulomb local-field term
which lifts the double degeneracy between the dark excitons
made of electronic states from the two valleys
(the possibility of such splitting was briefly mentioned in
\cite{Dery2015}).
This local-field effect is analogous to the exchange energy
shift of the $Z$~excitons in semiconductor quantum
wells~\cite{Andreani1990,Chen1988}, and produces an energy
splitting which we very roughly estimate as about 10~meV.
Its precise evaluation requires a microscopic treatment
on the atomic scales, similar to that in
\cite{Echeverry2016}.
Crucially, the whole oscillator strength of the interband
spin-flip dipole goes into the higher-energy component.
Thus, we find that the dark intravalley $A$~exciton has,
in fact, two components, one which is {truly} dark and
the other one about 100--1000 times darker than the bright
exciton, so it can be called ``dim''.

Although our results for the dark exciton splitting and
decay rates apply both to molybdenum- and tungsten-based
compounds (the discussed decay mechanisms work quite
analogously in the two cases), in the context of
photoluminescence (PL) our results are more relevant to
the WX$_2$ case, where the dark excitons have lower
energies and thus are more populated at low temperatures
than the bright ones.
Still, we are not yet in the position to
unambiguously identify various peaks observed in the
low-temperature PL spectra of WSe$_2$~%
\cite{Jones2013,Wang2014,Arora2015,You2015,Mitioglu2015,%
Koperski2015,Srivastava2015,Smolenski2015}
as being due to dark or dim excitons. Our work focuses
on intravalley spin-flip processes only, and a detailed
theoretical study of various valley-flip processes is
still required to create a complete picture. Still, our
results give some indications for experimental studies
which could shed some light on the origin of various
features in the spectrum.
For example, because of the dark-dim splitting, at low
temperatures the emission from the dim exciton should be
suppressed by an activation factor, analogously to that
from the bright one, which should manifest itself in the
temperature dependence of the relative peak intensities.
The transition dipole of the dim component being
perpendicular to the plane, it could be identified by
the angular distribution of the emission.
The dark component can be made decay by applying an
in-plane magnetic field, and the corresponding rate can
be made comparable to that of the dim component for a
sufficiently high field.

The paper is organized as follows.
In Sec.~\ref{sec:model} we describe the effective model for
electrons in the TMDC monolayer, the excitonic states, and specify
how different spin-flip mechanisms enter the model.
In Sec.~\ref{sec:polarization}, we discuss spin-mixing of excitonic
states and calculate the ``mechanical'' susceptibilities of the
TMDC monolayer, which determine the exciton coupling to the
macroscopic electromagnetic field.
In Sec.~\ref{sec:radiative} we study the effect of the exciton
coupling to the electromagnetic field and compute the radiative
energy shifts and decay rates for the excitons.
Finally, in Sec.~\ref{sec:conclusions} we summarize our results
and discuss some of their implications and perspectives.
Some details of calculations are presented in several appendices.

\section{The model}
\label{sec:model}

We set $\hbar=1$ throughout the paper,
and write the electronic Hamiltonian of a TMDC monolayer as
\begin{equation}\label{H=}
\hat{H}=\hat{H}_\mathrm{b}+\hat{H}_\mathrm{ee}
+\hat{H}_\mathrm{sf}.
\end{equation}
Its three terms will be discussed in the following subsections.

\subsection{Electronic bands}

We assume the monolayer to be in the $xy$ plane,
the three-dimensional position vector $\vec{R}$ represented
$\vec{R}\equiv(x,y,z)\equiv(\vec{r},z)$, so that $\vec{r}$
denotes the position in the monolayer plane.
The first term in Eq.~(\ref{H=}), $\hat{H}_\mathrm{b}$,
is the usual effective two-band
Hamiltonian~\cite{Kormanyos2015}
\begin{equation}
\hat{H}_\mathrm{b}=\sum_{\tau=\pm{1}}
\int\hat\psi^\dagger_\tau(\vec{r})\,
\mathcal{H}_\tau(-i\boldsymbol{\nabla})\,
\hat\psi_\tau(\vec{r})\, d^2\vec{r},
\end{equation}
written in terms of four-component column operators
$\hat\psi_\tau=[\hat\psi_{\tau,\mathrm{c},\uparrow}\;
\hat\psi_{\tau,\mathrm{c},\downarrow}\;
\hat\psi_{\tau,\mathrm{v},\uparrow}\;
\hat\psi_{\tau,\mathrm{v},\downarrow}]^T$,
where $\tau=\pm{1}$ labels the two valleys $\pm\vec{K}$,
and $\mathcal{H}_\tau(\vec{k})$ is the $4\times{4}$ matrix,
written in the block form,
\begin{equation}\label{Htauk=}
\mathcal{H}_\tau(\vec{k})=
\left[\begin{array}{cc}
E_\mathrm{g}+\tau\Delta_\mathrm{c}\sigma_z+\alpha_\mathrm{c}k^2 &
v(\tau k_x-i k_y) \\
v(\tau k_x+i k_y) &
\tau\Delta_\mathrm{v}\sigma_z-\alpha_\mathrm{v}k^2\end{array}\right].
\end{equation}
Here $\sigma_z$ is the third Pauli matrix in the spin subspace,
$E_\mathrm{g}$ is the band gap, $v$~is the velocity matrix
element between the band extrema Bloch functions (which can be
made real by an appropriate choice of the relative phase between
the conduction and valence band Bloch functions).
The coefficients $\alpha_\mathrm{c,v}$ are related to the
electron and hole effective masses $m_\mathrm{e,h}$ as
\begin{equation}\label{meh=}
\frac{1}{2m_\mathrm{e,h}}=\alpha_\mathrm{c,v}
+\frac{v^2}{E_\mathrm{g}}.
\end{equation}
Typically, $E_\mathrm{g}\sim{2}.5-3\:\mbox{eV}$,
$v\sim{2}.5\:\mbox{eV}\cdot\mbox{\AA}$,
$m_\mathrm{e}\sim{m}_\mathrm{h}\sim{0}.5\,m_0$ for MoX$_2$
and $\sim{0}.3\,m_0$ for WX$_2$ ($m_0$ being the free electron
mass).

$\Delta_\mathrm{c,v}$ in Eq.~(\ref{Htauk=}) is half of the spin
splitting in the conduction/valence bands.
We distinguish the $\pm\vec{K}$ valleys
by assuming the valence band states to originate predominantly
from the metallic orbitals with the $z$~projections of the angular
momentum equal to $\pm{2}$ at the $\pm\vec{K}$ point, respectively
\cite{Xiao2012,Liu2013,Mattheiss1973,Voss1999,Lebegue2009,Zhu2011,%
Kadantsev2012,Chang2013,Cappelluti2013,Song2013}.
This fixes $\Delta_\mathrm{v}>0$, as well as the valley-dependent
optical selection rules: the left/right circular polarisation can
be absorbed in the $\pm\vec{K}$ valley, respectively, as the
conduction band states originate mostly from zero-angular-momentum
metallic orbitals and the angular momentum is conserved modulo~3
due to the three-fold crystal rotation symmetry. Typically,
$\Delta_\mathrm{v}\sim{1}00-200\:\mbox{meV}$ in MoX$_2$ and
$400-500\:\mbox{meV}$ in WX$_2$. $\Delta_\mathrm{c}$ is much
weaker, usually a few tens of meV, with the exception of MoS$_2$
where it is extremely weak, $\sim{3}\:\mbox{meV}$.
The sign of $\Delta_\mathrm{c}$ depends on the material:
$\Delta_\mathrm{c}<0$ ($\Delta_\mathrm{c}>0$) in molybdenum-based
(tungsten-based) compounds. Because of this, the lowest-energy
interband transition in tungsten-based compounds involves either
a spin flip or valley switching.

\subsection{Coulomb interaction and excitons}
\label{ssec:Coulomb}

The second term in Eq.~(\ref{H=}), $\hat{H}_\mathrm{ee}$,
represents the Coulomb interaction between electrons,
\begin{align}
\label{Hee=}
&\hat{H}_\mathrm{ee}=\frac{1}2\int{V}(\vec{r}-\vec{r}')\,
\hat{\rho}(\vec{r})\,\hat{\rho}(\vec{r}')\,
d^2\vec{r}\,d^2\vec{r}',\\
&\hat\rho(\vec{r})=\sum_{\tau=\pm{1}}
\hat\psi^\dagger_\tau(\vec{r})
\left[\begin{array}{cc} 1 & 1 \\ 1 & 1 \end{array}\right]
\hat\psi_\tau(\vec{r}).
\label{rhomatrix=}
\end{align}
Here each element of the block $2\times{2}$ matrix should be
understood as the unit matrix in the spin space.
The diagonal
(intraband) blocks are responsible for direct electron-electron
and hole-hole repulsion, while the off-diagonal blocks encode
the direct elctron-hole attraction, which gives
rise to the excitonic bound state.
We included only the intra-valley part of the electron
density $\hat\rho(\vec{r})$, as it is the long-range part
of the Coulomb interaction (on the scale of lattice constant)
which is responsible for the exciton formation.

The wave function of
the electron-hole relative motion,
$\Phi(\vec{r}_\mathrm{e}-\vec{r}_\mathrm{h})$, is obtained
from the corresponding Schr\"odinger equation,
\begin{equation}\label{excSchrodinger=}
\left[-\frac{\nabla^2}{2m'}+V(\vec{r})\right]\Phi(\vec{r})=
E\,\Phi(\vec{r}),
\end{equation}
where
$m'=m_\mathrm{e}m_\mathrm{h}/(m_\mathrm{e}+m_\mathrm{h})$
is the reduced mass.
The lowest eigenvalue $E=-E_\mathrm{b}$ of
Eq.~(\ref{excSchrodinger=}) defines the exciton binding
energy~$E_\mathrm{b}$.
As we are working in the parabolic approximation for the
electronic dispersion and assume the electron and the hole
masses to be spin-independent, the wave function $\Phi(\vec{r})$
and the binding energy~$E_\mathrm{b}$ are the same for all spin
and valley configurations of the electron and the hole.

We do not specify the explicit form of the pair interaction
potential ${V}(\vec{r})$, as we will not be solving
Eq.~(\ref{excSchrodinger=}). It is known that because of strong
dielectric confinement in the TMDC monolayer, the interaction
potential is not $1/r$, so the bound state wave
functions do not have a hydrogenic form
\cite{Berkelbach2013,Qiu2013,Chernikov2014,Ye2014,He2014}.
Instead, we will assume $\Phi(\vec{r})$ and $E_\mathrm{b}$
to be known and treat them as input parameters. In fact, we will
not need the whole wave function $\Phi(\vec{r})$, but only its
value $\Phi(0)$ at the coinciding electron and hole positions.
By the order of magnitude, $\Phi(0)$ is the inverse radius of
the excitonic bound state. Estimating it as
$\Phi(0)\sim\sqrt{m'E_\mathrm{b}}$ with the binding energy
$E_\mathrm{b}\sim{0}.5-1\:\mbox{eV}$
\cite{Ramasubramaniam2012,Berkelbach2013,Qiu2013}, gives
$\Phi(0)\sim{0}.1-0.2\:\mbox{\AA}^{-1}$, while from the measured
diamagnetic shift of the exciton energy one infers values slightly
below ${0}.1\:\mbox{\AA}^{-1}$ \cite{Stier2016}.

The contribution from the off-diagonal matrix elements
in Eq.~(\ref{rhomatrix=}) corresponds to the exchange
interaction between the electron and the hole. It will be
taken into account in Sec.~\ref{sec:radiative} in the framework
of macroscopic electrodynamics.

\subsection{Spin-flip processes}

The third term in Eq.~(\ref{H=}), $\hat{H}_\mathrm{sf}$,
is the spin flip which arises when the reflection symmetry
in the~$z$ direction is broken~\cite{Bychkov1984}.
We write this term as~\cite{Ochoa2013}
\begin{align}\label{HRashba=}
&\hat{H}_\mathrm{sf}=-\mathcal{E}_z\sum_{\tau=\pm{1}}
\int\hat\psi^\dagger_\tau(\vec{r})\,\mathcal{D}_\tau\,
\hat\psi_\tau(\vec{r})\, d^2\vec{r},\\
&\mathcal{D}_\tau=\left[\begin{array}{cccc}
0 & 0 & 0 & -id_z\delta_{\tau,-1}\\
0 & 0 & -id_z\delta_{\tau,1} & 0 \\
0 & id_z\delta_{\tau,1} & 0 & 0 \\
id_z\delta_{\tau,-1} & 0 & 0 & 0 \end{array}\right].
\end{align}
We assumed that the reflection symmetry is broken by
an external perpendicular electric field $\mathcal{E}_z$,
introduced explicitly. Then, by definition,
the operator multiplying $-\mathcal{E}_z$ is nothing but
the $z$~component of the electric dipole moment operator.
If there is no external electric field, but the reflection
symmetry is broken by some other mechanism (e.~g., van der
Waals interaction with a dielectric substrate), the coupling
Hamiltonian still has the form~(\ref{HRashba=}) which is
fixed by the symmetry, but instead of $d_z\mathcal{E}_z$
its strength is determined by another parameter having the
dimensionality of energy. For example, the parameter
$\lambda_\mathrm{ext}$ from \cite{Ochoa2013} is related
to $d_z\mathcal{E}_z$ as
$d_z\mathcal{E}_z=2\lambda_\mathrm{ext}$.
The parameter~$d_z$ is real, which is fixed by the combination of time-reversal symmetry and the reflection symmetry $x\to-x$
(see Appendix~\ref{app:dipole_z_term} for details).
The peculiar valley structure of the matrix $\mathcal{D}_\tau$
arises because the total angular momentum
(orbital plus spin) is conserved modulo~3 due to the
crystal symmetry:
in $\vec{K}$ ($-\vec{K}$) valley only spin-up (spin-down)
states in the valence band can couple to only spin-down
(spin-up) in the conduction band, the corresponding total
angular momentum change being precisely $\pm{3}$.

\begin{figure}
\includegraphics[width=8cm]{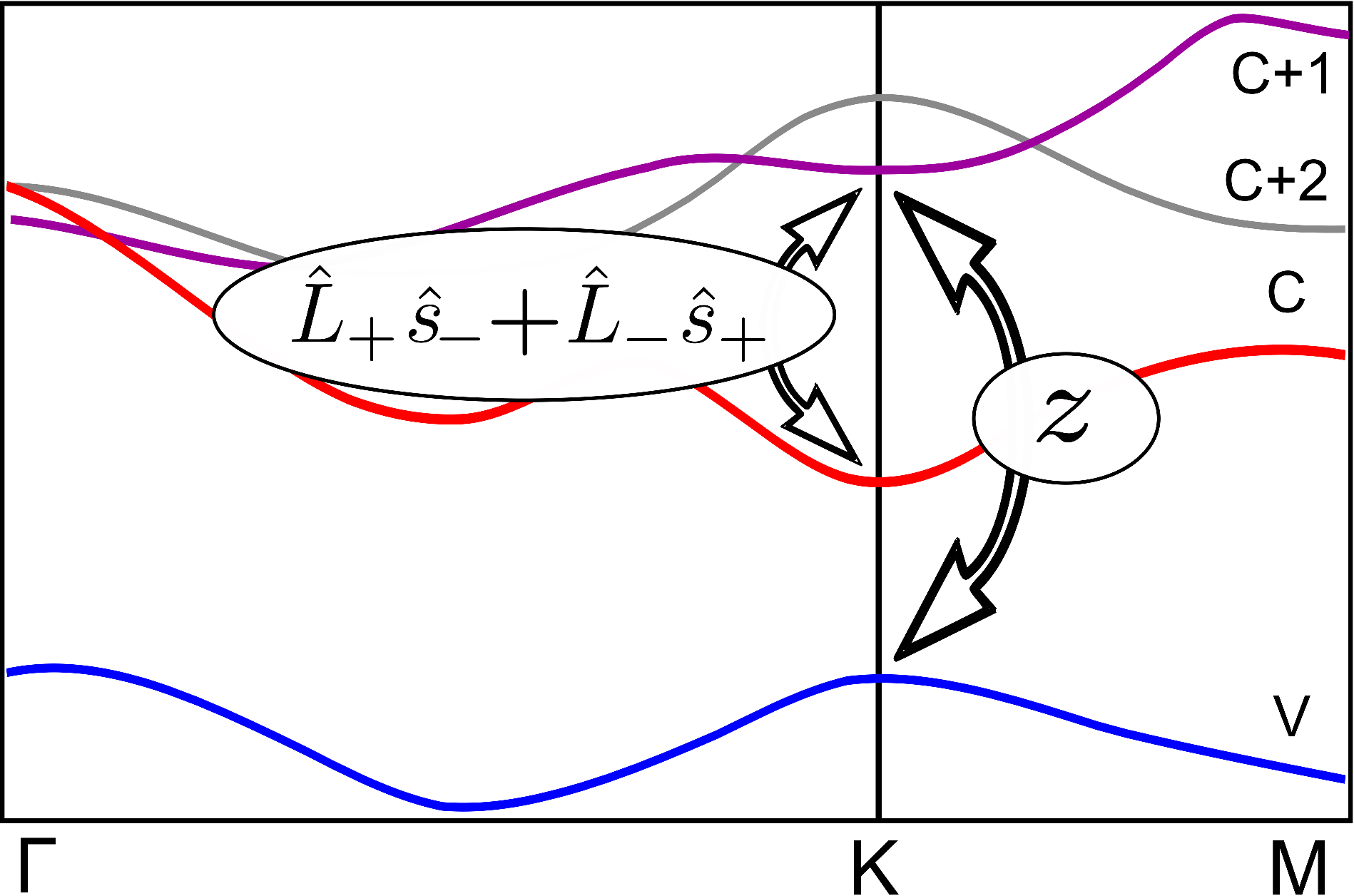}
\caption{\label{fig:spin_flip} 
A sketch of the TMDC monolayer band structure,
including the valence band (v) and three lowest conduction
bands (c, c$+1$, c$+2$).
The arrows represent the virtual transitions which lead
to the intraband dipole spin-flip transition, described by
Eq.~(\ref{HRashba=}).
The spin splitting of the bands is not shown for simplicity.
}
\end{figure}

The intraband dipole spin-flip Hamiltonian~(\ref{HRashba=})
can be understood as a combination of
the in-plane part $\hat{L}_+\hat{s}_-+\hat{L}_-\hat{s}_+$
of the atomic spin-orbit coupling
$\propto\hat{\vec{L}}\cdot\hat{\vec{s}}=\hat{L}_z\hat{s}_z
+(\hat{L}_+\hat{s}_-+\hat{L}_-\hat{s}_+)/2$ (here
$\hat{\vec{L}}$ and $\hat{\vec{s}}$ are the orbital
angular momentum and the spin of the electron,
$\hat{L}_\pm\equiv\hat{L}_x\pm i\hat{L}_y$,
$\hat{s}_\pm\equiv\hat{s}_x\pm i\hat{s}_y$)
and of the electric potential $-e\mathcal{E}_zz$ which
acts as a perturbation on the microscopic wave functions
of the electron in the crystal~\cite{Kormanyos2014}.
For example, under the action of the
$\hat{L}_+\hat{s}_-$ part of the atomic spin-orbit
coupling, a spin-up electron in the conduction band
can first perform a virtual transition to the next
conduction band (denoted here by c$+1$), which flips
its spin. It is crucial that the band c$+1$ is odd
under the reflection $z\to{-}z$, since the conduction
band is even, and the operators $\hat{L}_\pm$ are odd,
so the matrix elements
$\langle\mathrm{c}+1|\hat{L}_\pm|\mathrm{c}\rangle$
are allowed.
Next, the electron performs the transition to the
valence band, induced by the perturbation
$-e\mathcal{E}_zz$. Again, because the valence band
is even and $z$ is odd, the matrix element
$\langle\mathrm{v}|z|\mathrm{c}+1\rangle\neq{0}$.
The result of these two virtual processes is the
transition from the conduction to the valence band,
accompanied by the spin flip, as shown schematically in
Fig.~\ref{fig:spin_flip}. Such transition is forbidden
if there is no perturbation to break the parity
$z\to-z$.
Combining the interband Hamiltonian $\hat{H}_\mathrm{sf}$
with the off-diagonal terms in Eq.~(\ref{Htauk=}) in
the first-order perturbation
theory, one can project the spin-flip terms on the conduction
and valence bands (see Appendix~\ref{app:diag_hamiltonian}).
Upon projection, they acquire the familiar Rashba-type
form~\cite{Bychkov1984}, linear in momentum~$\vec{k}$:
\begin{align}
\label{Rcvb=}
\mathcal{H}^\mathrm{R}_\tau(\vec{k})=
\frac{v\mathcal{E}_z}{E_\mathrm{g}}d_z
[\vec{k}\times\boldsymbol{\sigma}]_z
\left[\begin{array}{cc}
1 & 0 \\ 0 & -1
\end{array}\right].
\end{align}
This form matches the one studied in \cite{Kormanyos2014}
where the Rashba coupling strength was parametrized by
$\lambda_\mathrm{BR}$, related to $d_z\mathcal{E}_z$ as
$i d_z\mathcal{E}_z=\lambda_\mathrm{BR}E_\mathrm{g}/v$.
The parameter $|\lambda_\mathrm{BR}|$ was estimated in
\cite{Kormanyos2014} for MoS$_2$, MoSe$_2$, WS$_2$,
WSe$_2$ monolayers suspended in vacuum and subject to an
external perpendicular electric field. This enables us
to extract
$|d_z/e|$ of about ${0}.02\:\mbox{\AA}$, $0.03\:\mbox{\AA}$,
$0.06\:\mbox{\AA}$, and $0.08\:\mbox{\AA}$, for
MoS$_2$, MoSe$_2$, WS$_2$, and WSe$_2$, respectively.

We are not aware of any estimates in the literature for
the Rashba coupling strength induced by a dielectric
substrate.
Therefore, we make our own crude estimate of the effective
electric field in Appendix~\ref{app:substrate}, which gives
$\mathcal{E}_z\sim{0.1}\:\mbox{V/\AA}$
for a WX$_2$ monolayer lying on a glass-like substrate with a
dielectric constant $\sim{4}$ and vacuum above.

Finally, we note that in addition to Rashba effect in a
perpendicular electric field, the spin can be flipped by
the Zeeman effect if an in-plane magnetic field $\vec{B}_\|$
is applied.
This effect is taken into account straightforwardly by adding
$g_\mathrm{c,v}\mu_B\vec{B}_\|\cdot\boldsymbol{\sigma}/2$
to the diagonal matrix elements in Eq.~(\ref{Htauk=}),
where $\mu_B\approx{58}\:\mu\mbox{eV/T}$ is the Bohr magneton
and $g_\mathrm{c,v}\approx-2$ is the
in-plane gyromagnetic ratio for electrons in the conduction
and valence band.

\subsection{Coupling to photons}

The Hamiltonian of interaction between the electrons and the
transverse photons, which are described by the long-wavelength
vector potential $\vec{A}(\vec{r})$, is usually obtained from
the requirement of gauge invariance by performing the Peierls
substitution in Eq.~(\ref{Htauk=}):
$\mathcal{H}_\tau(\vec{k})\to
\mathcal{H}_\tau(-i\boldsymbol{\nabla}-(e/c)\vec{A})$.
However,
besides coupling to photons via gauge potentials, electrons
can couple directly to electric and magnetic fields. Such
terms are gauge invariant and cannot be deduced from the bare
electronic Hamiltonian in the envelope function approximation,
as they correspond to the effect of electromagnetic field
on the microscopic Bloch functions.

In the previous subsection, we have already presented such
terms, corresponding to Rashba and Zeeman effects.
It is crucial for our analysis that, once the dipole operator
in Eq.~(\ref{HRashba=}) is determined, it can also describe
the electronic coupling to the photon field. Namely,
$\mathcal{E}_z$ can be understood not just as a static
external field, but as the electric field of photons as well.
In this case, one should write
$\mathcal{E}_z=\mathcal{E}_z(\vec{r})$ and put it inside
the space integral.
The corresponding Hamiltonian can then describe interband
photon absorption/emission, accompanied by the spin flip
(the existence of such coupling was briefly mentioned in
\cite{Glazov2014}).

\section{Polarization susceptibility}
\label{sec:polarization}

\subsection{General definitions}

The main quantity which we use to describe the interaction of
excitons with light is the susceptibility
$\chi_{ij}(\vec{q},\omega)$, defined as the mechanical response
of the excitonic polarization to the electric field at the
frequency $\omega$ and the in-plane wave vector~$\vec{q}$.
(The term ``mechanical'' implies that only the direct Coulomb
interaction is taken into account for the moment; the exchange
interaction will be included in the framework of macroscopic
electrodynamics.)
We define the polarization $\vec{P}$ as the dipole moment per
unit area. It is a three-dimensional vector, whose $z$ component
is defined via~(\ref{HRashba=}) as the operator multiplying
$-\mathcal{E}_z$,
while the in-plane components
are most conveniently defined via the in-plane current
$\vec{j}=\partial\vec{P}/\partial{t}$. The current operator is
obtained in the usual way from
$\partial\mathcal{H}_\tau(\vec{k})/\partial\vec{k}$, where we
retain only the leading terms at $k\to{0}$:
\begin{equation}
\label{current=}
\hat{\vec{j}}(\vec{r})=ev\sum_\tau\hat\psi^\dagger_\tau(\vec{r})
\left[\!\!\begin{array}{cc} 0 \!& \tau\vec{e}_x- i\vec{e}_y \\
\tau\vec{e}_x+ i\vec{e}_y \!& 0 \end{array}\!\!\right]
\hat\psi_\tau(\vec{r}),
\end{equation}
where $\vec{e}_x,\vec{e}_y$ are the unit vectors in the
corresponding directions.
Using the Kubo formula, whose general form for the
susceptibility $\chi_{\mathcal{AB}}(\omega)$ determining
the response of a quantity $\mathcal{A}$ to the periodic
force which couples to a quantity $\mathcal{B}$ reads as
\[
\chi_{\mathcal{AB}}(\omega)=i\int\limits_0^\infty
\left\langle[\hat{\mathcal{A}}(t),\hat{\mathcal{B}}(0)]\right\rangle
e^{i\omega{t}}\,dt,
\]
where $\hat{\mathcal{A}}(t)$, $\hat{\mathcal{B}}(0)$ are
the time-dependent operators in the Heisenberg representation
of the unperturbed Hamiltonian and the average is taken over
the unperturbed equilibrium state, we obtain the following
expression for the polarization susceptibility:
\begin{align}
\label{chidef=}
&\chi_{\alpha\beta}(\vec{r},\vec{r}',\omega)=
\sum_\nu\frac{\langle{0}|\hat{j}_\alpha(\vec{r})|\nu\rangle
\langle\nu|\hat{j}_\beta(\vec{r}')|0\rangle}%
{E_\nu^2(E_\nu-\omega-i0^+)},\\
&\chi_{\alpha{z}}(\vec{r},\vec{r}',\omega)=i
\sum_\nu\frac{\langle{0}|\hat{j}_\alpha(\vec{r})|\nu\rangle
\langle\nu|\hat{P}_z(\vec{r}')|0\rangle}%
{E_\nu(E_\nu-\omega-i0^+)},\\
&\chi_{zz}(\vec{r},\vec{r}',\omega)=
\sum_\nu\frac{\langle{0}|\hat{P}_z(\vec{r})|\nu\rangle
\langle\nu|\hat{P}_z(\vec{r}')|0\rangle}{E_\nu-\omega-i0^+}.
\label{chidef_3=}
\end{align}
Here $|0\rangle$ and $|\nu\rangle$ are the ground and excited
electronic states, $E_\nu$ is their energy difference (that is,
the energy of the excitation~$\nu$), the indices
$\alpha,\beta=x,y$, and we omitted the non-resonant terms.
The infinitesimal imaginary part $i0^+$ in the denominators
reflects the causal nature of the susceptibility and determines
its analytical properties (the susceptibility must be analytical
in the upper complex half-plane of~$\omega$).
In the translationally invariant case, the susceptibilities
depend only on the difference $\vec{r}-\vec{r}'$, so
\begin{equation}
\chi_{ij}(\vec{r}-\vec{r}',\omega)=\int\frac{d^2\vec{q}}{(2\pi)^2}\,
e^{i\vec{q}\cdot(\vec{r}-\vec{r}')}\chi_{ij}(\vec{q},\omega).
\end{equation}
Here the in-plane components $x,y$ are labeled by the Greek
indices $\alpha,\beta$, while the Latin indices run over all
three dimensions, $i,j=x,y,z$.

In the following, we evaluate $\chi_{ij}(\vec{q},\omega)$ using
the eigenstates $|\nu\rangle$ which are the lowest-energy states
of the intravalley excitons, whose spins are mixed by the Rashba
coupling. The effect of the Zeeman coupling is treated analogously,
the result is summarized in the end of
Sec.~\ref{ssec:susceptibility}.

\subsection{Excitonic states}
\label{ssec:excitons}

We start by constructing the zero-approximation states, i.~e.,
those in the absence of the Rashba coupling.
First, each intravalley exciton can be characterized by the
index~$\tau$, indicating the valley in which the electronic
transition takes place.
It gives rise to two species of intravalley excitons, which
can be distinguished by the valley index $\tau=\pm{1}$. Further,
in each valley the conduction and valence bands are spin split,
so the excitonic states can be labeled by a pair of spin indices
$s_\mathrm{c},s_\mathrm{v}\!\!\!=\,\uparrow,\downarrow$ referring to
the conduction and the valence band, respectively (the hole spin
is given by $-s_\mathrm{v}$).
Finally, the excitonic state is characterized by a center-of-mass
momentum $\vec{q}$. So, the zero-approximation excitonic states
are written as
\begin{align}
|\tau,s_\mathrm{c},s_\mathrm{v},\vec{q}\rangle_0=
\int& d^2\vec{r}_\mathrm{e}\,d^2\vec{r}_\mathrm{h}\,
\frac{e^{i\vec{q}\vec{r}_\mathrm{cm}}}{\sqrt{S}}\,
\Phi(\vec{r}_\mathrm{e}-\vec{r}_\mathrm{h})
\times \nonumber \\& \times
\hat{\psi}_{\tau,\mathrm{c},s_\mathrm{e}}^\dag(\vec{r}_\mathrm{e})\,
\hat{\psi}_{\tau,\mathrm{v},s_\mathrm{v}}(\vec{r}_\mathrm{h})
|0\rangle.
\label{state0=}
\end{align}
Here, $S$ is the sample area,
$\vec{r}_\mathrm{cm}$ is the center-of-mass coordinate,
$\vec{r}_\mathrm{cm}\equiv(m_\mathrm{e}\vec{r}_\mathrm{e}+
m_\mathrm{h}\vec{r}_\mathrm{h})/(m_\mathrm{e}+m_\mathrm{h})$,
and $\Phi(\vec{r}_\mathrm{e}-\vec{r}_\mathrm{h})$ is the
normalized wave function of the relative electron-hole motion,
corresponding to the lowest bound state, which was discussed
in Sec.~\ref{ssec:Coulomb}.
States~(\ref{state0=}) are normalized as
\begin{equation}
{}_0\langle\tau,s_\mathrm{c},s_\mathrm{v},\vec{q}|
\tau',s_\mathrm{c}',s_\mathrm{v}',\vec{q}'\rangle_0=
\delta_{\tau\tau'}\delta_{s_\mathrm{c}s_\mathrm{c}'}
\delta_{s_\mathrm{v}s_\mathrm{v}'}\delta_{\vec{q}\vec{q}'},
\end{equation}
and have the energies
\begin{equation}
E_{\tau,s_\mathrm{c},s_\mathrm{v}}^{(0)}(\vec{q})=
E_\mathrm{g}-E_\mathrm{b}
+\tau(s_\mathrm{c}\Delta_{\mathrm{c}}-s_\mathrm{v}\Delta_{\mathrm{v}})
+\frac{q^2}{2m_\mathrm{ex}},
\end{equation}
where we introduced the excitonic mass
$m_\mathrm{ex}\equiv{m}_\mathrm{e}+m_\mathrm{h}$.

The Rashba coupling flips electron and hole spins and mixes
the excitonic states with different $s_\mathrm{c},s_\mathrm{v}$.
As $\Delta_\mathrm{c}/\Delta_\mathrm{v}\sim{0.1}\ll{1}$, we
neglect the spin-flip in the valence band, and consider it only
in the conduction band. It can be described by an effective
$2\times{2}$ Hamiltonian in the basis of
\{$|\tau,\uparrow,s_\mathrm{v},\vec{q}\rangle_0$,
$|\tau,\downarrow,s_\mathrm{v},\vec{q}\rangle_0$\}, obtained by
the projection of the electronic Rashba Hamiltonian~(\ref{Rcvb=})
on the excitonic states:
\begin{equation}\label{Rashbaex=}
\hat{H}_{\tau,s_\mathrm{v}}^\mathrm{ex}(\vec{q})=
\left[\begin{array}{cc}
E_{\tau,\uparrow,s_\mathrm{v}}(\vec{q}) &  -i\lambda_\mathrm{ex}q_-  \\
i\lambda_\mathrm{ex}q_+ & E_{\tau,\downarrow,s_\mathrm{v}}(\vec{q})
\end{array}\right],
\end{equation}
where the excitonic Rashba coupling constant is given by
\begin{equation}\label{lambdaex=}
\lambda_\mathrm{ex}=\frac{vd_z\mathcal{E}_z}{E_\mathrm{g}}\,
\frac{m_\mathrm{e}}{m_\mathrm{e}+m_\mathrm{h}}.
\end{equation}
Description of spin flip in terms of the purely excitonic
effective Hamiltonian~(\ref{Rashbaex=})
assumes that the spin flip does not disturb the electron-hole relative
motion inside the exciton, which is guaranteed by the condition
$\Delta_c,\lambda_\mathrm{ex}q\ll{E}_\mathrm{b}$~\cite{footnote_1}.

The $2\times{2}$ Hamiltonian (\ref{Rashbaex=}) can be diagonalized
exactly; however, we are interested only in small momenta~$q$, so
perturbative expressions will suffice for us.
Up to second order in $\lambda_\mathrm{ex}$, the mixed-spin eigenstates
are given by
\begin{align}
\label{mixedspin=}
&|\tau,\uparrow,s_\mathrm{v},\vec{q}\rangle_\mathrm{mix}\!=\!
\frac{|\tau,\uparrow, s_\mathrm{v},\vec{q}\rangle_0}{1+w_\vec{q}/2} +
\frac{i\lambda_\mathrm{ex}q_+}{2\tau\Delta_\mathrm{c}}
|\tau,\downarrow,s_\mathrm{v},\vec{q}\rangle_0,\\
\label{mixedspin_2=}
&|\tau,\downarrow, s_\mathrm{v},\vec{q}\rangle_\mathrm{mix}\!=\!
\frac{|\tau,\downarrow, s_\mathrm{v},\vec{q}\rangle_0}{1+w_\vec{q}/2}
+\frac{i\lambda_\mathrm{ex} q_-}{2\tau\Delta_\mathrm{c}}
|\tau,\uparrow,s_\mathrm{v},\vec{q}\rangle_0,
\end{align}
where $w_\vec{q}\equiv\lambda_\mathrm{ex}^2q^2/(4\Delta_c^2)$.
To the same order, their energies are given by
\begin{equation}
E^\mathrm{mix}_{\tau,s_\mathrm{c},s_\mathrm{v}}(\vec{q})=
E_{\tau,s_\mathrm{c},s_\mathrm{v}}^{(0)}(\vec{q})
+\tau s_\mathrm{c}\,\frac{\lambda_\mathrm{ex}^2}{2\Delta_\mathrm{c}}\,q^2.
\end{equation}
The states with $s_\mathrm{c}=s_\mathrm{v}=\uparrow$ from the
$\vec{K}$ valley and those with
$s_\mathrm{c}=s_\mathrm{v}=\downarrow$ from the $-\vec{K}$ valley
are called the bright $A$ excitons. Those with the opposite
$s_\mathrm{c},s_\mathrm{v}$ have a higher energy (due to the
$\Delta_\mathrm{v}$ term) and are called the bright $B$ excitons.
Their energies are
\begin{align}
E_{A,B}^\mathrm{b}(\vec{q})\!=\!E_\mathrm{g}\!-\!E_\mathrm{b}
+\!\frac{q^2}{2m_\mathrm{ex}}\mp\Delta_\mathrm{v} 
\pm \Delta_\mathrm{c}\pm
\frac{\lambda_\mathrm{ex}^2}{2\Delta_\mathrm{c}}\,q^2.
\end{align}
The states obtained by flipping the spin in the conduction band
can be called the dark $A$ and $B$ excitons, their energies are
\cite{footnote_2}
\begin{equation}
E_{A,B}^\mathrm{d}(\vec{q})=E_{A,B}^\mathrm{b}(\vec{q})
\mp{2}\left(\Delta_\mathrm{c}
+\frac{\lambda_\mathrm{ex}^2}{2\Delta_\mathrm{c}}\,q^2\right).
\end{equation}

\subsection{Calculation of susceptibility}
\label{ssec:susceptibility}

We calculate the susceptibility $\chi_{ij}(\vec{q},\omega)$
from Eqs.~(\ref{chidef=})--(\ref{chidef_3=}) using the matrix elements
$\langle{0}|\hat{j}_\alpha(\vec{r})|\nu\rangle$ and
$\langle{0}|\hat{P}_z(\vec{r})|\nu\rangle$, which, in turn, are
obtained from expressions~(\ref{mixedspin=}) and (\ref{mixedspin_2=}) for the mixed-spin
eigenstates~$|\nu\rangle$ and from the matrix elements between
the zero-approximation states~(\ref{state0=}) following
from the definitions of the in-plane current and the $z$-dipole
moment operators [Eqs.~(\ref{current=}) and~(\ref{HRashba=})].
Namely, for the in-plane current we have
\begin{equation}
\langle0|\hat{j}_\alpha(\vec{r})|
\tau,s_\mathrm{c},s_\mathrm{v},\vec{q}\rangle_0
=ev\Phi(0)\frac{e^{i\vec{q}\vec{r}}}{\sqrt{S}}
\left(\tau\delta_{\alpha{x}}+i\delta_{\alpha{y}}\right)\delta_{s_\mathrm{c}s_\mathrm{v}}.
\end{equation}
The nonzero matrix elements of the $z$-dipole moment operator
between the ground state and the zero-approximation states are
\begin{align}
&&\langle{0}|\hat{P}_z(\vec{r})|1,\downarrow,\uparrow,\vec{q}\rangle_0=
id_z\,\Phi(0)\,\frac{e^{i\vec{q}\vec{r}}}{\sqrt{S}},\\
&&\langle{0}|\hat{P}_z(\vec{r})|{-1},\uparrow,\downarrow,\vec{q}\rangle_0=
id_z\,\Phi(0)\,\frac{e^{i\vec{q}\vec{r}}}{\sqrt{S}}.
\end{align}
Summation over $\tau=\pm1$ in combination with the independence
of the energies $E_{A,B}^\mathrm{b,d}(\vec{q})$ on the valley
index~$\tau$ restores the in-plane isotropy in the final
expressions for the susceptibility, written to the second order
in~$d_z$:
\begin{align}
\label{chi=}
&\chi_{\alpha\beta}(\vec{q},\omega)=2\delta_{\alpha\beta}
\frac{[ev\Phi(0)]^2}{E_A^2}
\left[\frac{1-w_\vec{q}}{E_A^\mathrm{b}(\vec{q})-\omega}
+\frac{w_\vec{q}}{E_A^\mathrm{d}(\vec{q})-\omega}\right] \nonumber \\
&\hspace*{1.5cm}{}+2\delta_{\alpha\beta}
\frac{[ev\Phi(0)]^2}{E_B^2}\left[
\frac{1-w_\vec{q}}{E_B^\mathrm{b}(\vec{q})-\omega}
+\frac{w_\vec{q}}{E_B^\mathrm{d}(\vec{q})-\omega}\right],\\
\label{chinplane=}
&\chi_{{z}\alpha}(\vec{q},\omega)=-\chi_{\alpha{z}}(\vec{q},\omega)=
\nonumber \\ &\hspace*{.8cm}{}{}
=i\,
\frac{q_\alpha\lambda_\mathrm{ex}}{\Delta_c}d_z\frac{ev\Phi^2(0)}{E_A}
\left[\frac{1}{E_A^\mathrm{b}(\vec{q})-\omega}
-\frac{1}{E_A^\mathrm{d}(\vec{q})-\omega}\right],
\\
&\chi_{zz}(\vec{q},\omega)=2d_z^2\Phi^2(0)\,
\frac{1}{E_A^\mathrm{d}(\vec{q})-\omega}.
\label{chioutofplane=}
\end{align}
Here we omitted the infinitesimal imaginary part in the denominator
to keep the formulas more compact and approximated
$E_{A,B}^\mathrm{b,d}(\vec{q})\approx{E}_{A,B}$ in the
non-resonant prefactor, neglecting the dispersion and the
conduction band splitting. Note also that
$\Phi(0)$ is a real quantity.

From Eq.~(\ref{chi=}), we see the crucial role played by the
quantity
$w_\vec{q}\equiv\lambda_\mathrm{ex}^2q^2/(4\Delta_c^2)$,
introduced in Sec.~\ref{ssec:excitons}.
It represents the spectral weight transferred from the bright
to the dark excitons by the Rashba coupling. Let us estimate
its order of magnitude. Taking the numerical values typical
of WSe$_2$, $cq=1.7\:\mbox{eV}$, the experimentally
determined splitting
$2\Delta_\mathrm{c}=-30\:\mbox{meV}$~\cite{Zhang2015},
and $\lambda_\mathrm{ex}=9\:\mbox{meV}\cdot\mbox{\AA}$ from
\cite{Kormanyos2014} for a quite strong perpendicular
electric field $\mathcal{E}_z=0.1\:\mbox{V/\AA}$, we still
obtain a very small value of
$w_\vec{q}\approx{0}.7\times{10}^{-7}$.

If instead of the perpendicular electric field~$\mathcal{E}_z$,
an in-plane magnetic field~$B_\|$ is applied, spin mixing due
to the Zeeman effect can be taken into account by full analogy
with the Rashba mixing. In fact, it is sufficient to
replace $\lambda_\mathrm{ex}q\to|g_\mathrm{c}|\mu_BB_\|/2$,
so Eq.~(\ref{chinplane=}) has the same form, but the transferred
spectral weight
$w_\vec{q}=(g_\mathrm{c}\mu_BB_\|)^2/(16\Delta_\mathrm{c}^2)$.
For a magnetic field $B_\|=10\:\mbox{T}$, we obtain a numerical
estimate $w_\vec{q}\approx{4}\times{10}^{-4}$.
The in-plane direction of the off-diagonal component
$\chi_{\alpha{z}}$ is determined not by $\vec{q}$, but by the
magnetic field, $\chi_{\alpha{z}}\propto{B}_{\|\alpha}$.

\section{Exciton radiative shifts and decay rates}
\label{sec:radiative}

\subsection{General scheme}

Here we consider the interaction of excitons in the TMDC
monolayer
with the electromagnetic field. The monolayer is assumed to be
sandwiched between two semi-infinite media with dielectric
constants $\varepsilon_1$ and $\varepsilon_2$ occupying the
half-spaces with $z>0$ and $z<0$, respectively.
(Note that the values of $\vep_{1,2}$ at optical frequencies
$\sim{E}_A$ should be taken.) The free field
in such a structure is fully characterized by the
Green's function $D_{ij}(z,z';\vec{q},\omega)$,
which represents the response of the electric field,
$\mathcal{E}_i(z)\,e^{i\vec{q}\vec{r}-i\omega{t}}$,
to an external oscillating polarization
$P_j\delta(z-z')\,e^{i\vec{q}\vec{r}-i\omega{t}}$,
located in the plane $z=z'$. In the quantum
theory, this Green's function represents the retarded propagator
of the electric field; at the same time, it can be found from
the classical Maxwell equations~\cite{AGD}.
The radiative self-energy for the excitons at $z=0$ is
proportional to $D_{ij}(0,0;\vec{q},\omega)$, for which we
introduce the short-hand notation $\bar{D}_{ij}(\vec{q},\omega)$
[more precisely, it represents the projection on the spatial
profile of the excitonic polarization in the $z$-direction,
here assumed to be just $\delta(z)$].

The long-range exchange part of the Coulomb interaction shifts
the exciton energies and lifts the valley degeneracy. Exciton
coupling to the photons also shifts the excitonic frequencies
and leads to the radiative decay. All these effects can be
described by studying the linear system
\begin{equation}\label{Psystem=}
P_i=\chi_{ij}(\vec{q},\omega)\,\bar{D}_{jk}(\vec{q},\omega)\,P_k.
\end{equation}
For each~$\vec{q}$, it has non-trivial solutions for some
complex values of~$\omega$ whose real parts give the shifted
exciton energies, and the imaginary parts (with the opposite
sign and multiplied by~2)
represent the radiative decay rates. This procedure is equivalent
to finding the poles of the full layer susceptibility (i.~e.,
dressed by the exchange interaction and coupling to photons)
in the complex plane of~$\omega$, or to finding the poles of
the monolayer reflectivity.

Calculation of $\bar{D}_{ij}(\vec{q},\omega)$ from the Maxwell
equations is quite standard and is given in
Appendix~\ref{app:Maxwell}. The result is
\begin{align}
\bar{D}_{\alpha\beta}(\vec{q},\omega)={}&{}
\left(\delta_{\alpha\beta}-\frac{q_\alpha{q}_\beta}{q^2}\right)
\frac{4\pi i\omega^2/c^2}{q_{1z}+q_{2z}}+{}\nonumber\\
{}&{}+\frac{q_\alpha{q}_\beta}{q^2}\,
\frac{4\pi i(q_{1z}/\vep_1)(q_{2z}/\vep_2)}{q_{1z}/\vep_1+q_{2z}/\vep_2},
\label{Dab=}\\
\bar{D}_{{z}\alpha}(\vec{q},\omega)={}&
-\bar{D}_{\alpha{z}}(\vec{q},\omega)=\nonumber\\
&=2\pi iq_\alpha\,
\frac{q_{1z}/\vep_1-q_{2z}/\vep_2}{q_{1z}/\vep_1+q_{2z}/\vep_2},
\label{Daz=}\\
\bar{D}_{zz}(\vec{q},\omega)={}&{}
\frac{4\pi iq^2}{q_{1z}/\vep_1+q_{2z}/\vep_2}-2\pi\kappa_0.
\label{Dzz=}
\end{align}
Here $q_{1z}$ and $q_{2z}$ are the $z$~components of the
three-dimensional wave vector in the corresponding media:
\begin{equation}
q_{1z,2z}(\vec{q},\omega)=\sqrt{\vep_{1,2}(\omega+i0^+)^2/c^2-q^2},
\end{equation}
where the infinitesimal imaginary part
fixes the rule for the analytical continuation of the square
root in the upper complex half-plane of~$\omega$, inherited
from the analyticity of the response function
$\bar{D}_{ij}(\vec{q},\omega)$ in the upper half-plane.
This prescribes $\Im{q}_{1z,2z}>0$ for real $\omega$ in the
interval $|\omega|<cq/\sqrt{\vep_{1,2}}$, which corresponds
to evanescent waves.
The parameter $\kappa_0$ in Eq.~(\ref{Dzz=}) represents the electric
field of a double layer, arising from the excitonic polarization
in the $z$~direction (Appendix~\ref{app:Maxwell}), which is singular
in the limit of an infinitely thin layer, and cannot be determined
in the macroscopic framework, used here.
For excitons in semiconductor quantum wells of a sizable width, this
local-field effect could be treated properly in the envelope-function
approximation~\cite{Chen1988,Andreani1990}. In the atomically-thin
TMDC monolayer, it requires the full microscopic treatment of the
short-range exchange interaction, such as that in
\cite{Echeverry2016}. The consequences of this local-field
effect will be discussed  in the next subsection.

The tensor structure of $\chi_{ij}(\vec{q},\omega)$ and
$\bar{D}_{jk}(\vec{q},\omega)$, following from the in-plane isotropy
of the problem, determines how the two-fold valley degeneracy of the
mechanical excitons is lifted. The first family of solutions of
Eq.~(\ref{Psystem=}) is characterized by $\vec{P}$ lying in the
$xy$ plane, perpendicular to~$\vec{q}$. These transverse excitons
emit $s$-polarized light (transverse-electric, or TE modes).
The second family of solutions has $\vec{P}$ in the plane formed
by the vectors $\vec{q}$ and $\vec{e}_z$; its precise direction
is determined by the relative magnitude of different components
of $\chi_{ij}(\vec{q},\omega)$. These are longitudinal excitons,
which emit $p$-polarized light (transverse-magnetic, or TM modes).
Each of the two linear polarizations represents a linear
combination of the two circularly polarized excitons from each
valley with equal probability weights.
\begin{figure}
\includegraphics[width=8cm]{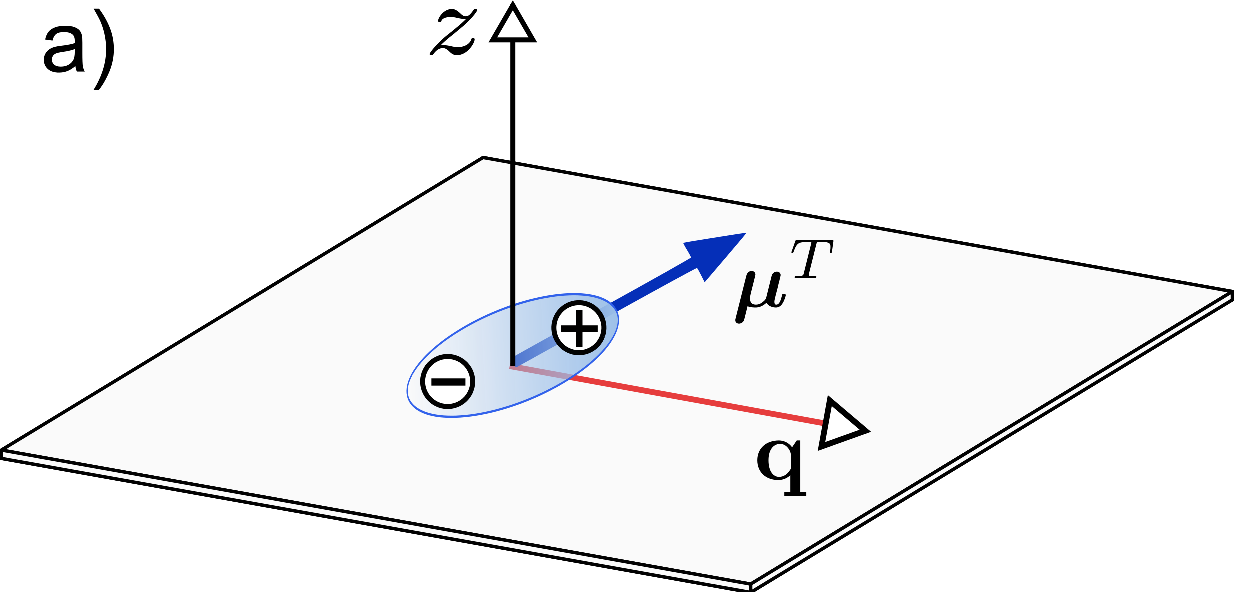}
\includegraphics[width=8cm]{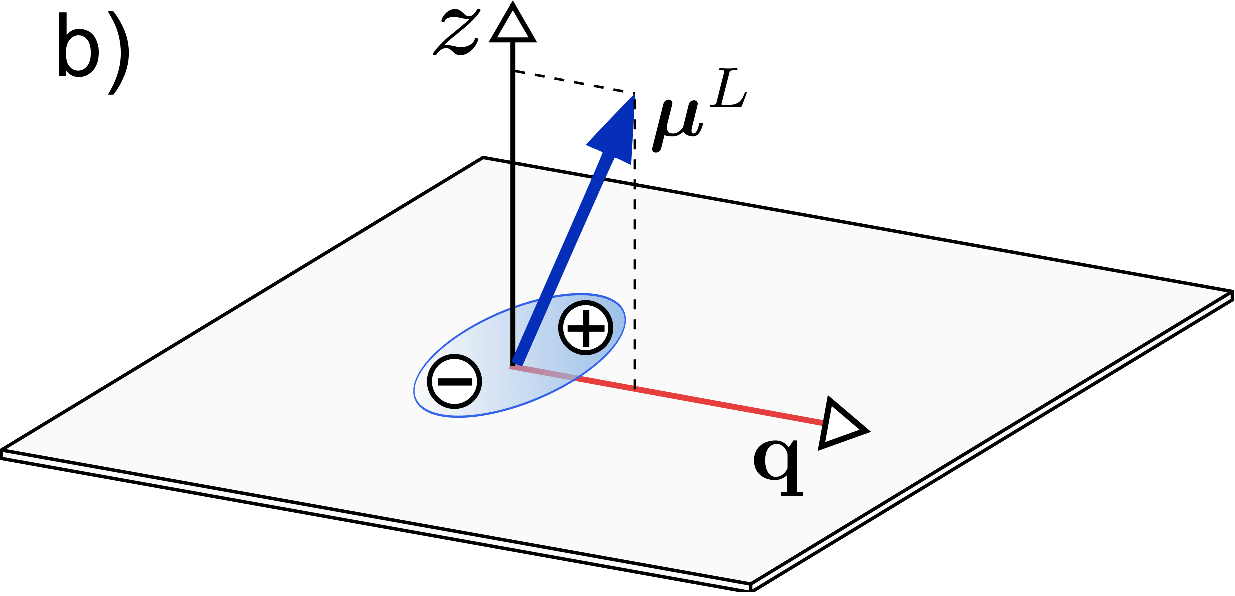}
\caption{\label{fig:LTexcitons} 
Orientation of the transition dipole moments
$\boldsymbol{\mu}_\nu^T$, $\boldsymbol{\mu}_\nu^L$ of the
transverse~(a) and longitudinal~(b) excitons with respect to the
crystal plane, the $z$ axis and the exciton momentum~$\vec{q}$.
}
\end{figure}
In the following, we treat Eq.~(\ref{Psystem=}) perturbatively.
Namely, we assume $\omega$ to be close to one of the poles $E_\nu$
of $\chi_{ij}(\vec{q},\omega)$. Near the pole, it can be
represented as
\begin{equation}\label{chipole=}
\chi_{ij}(\vec{q},\omega)\approx
\frac{\mu_{\nu{i}}^T(\mu_{\nu{j}}^T)^*
+\mu_{\nu{i}}^L(\mu_{\nu{j}}^L)^*}{E_\nu-\omega},
\end{equation}
with some vectors $\boldsymbol{\mu}_\nu^T$,
$\boldsymbol{\mu}_\nu^L$, which describe the
polarization of the transverse and longitudinal excitons.
The relative orientation of $\boldsymbol{\mu}_\nu^T$,
$\boldsymbol{\mu}_\nu^L$ with respect to the exciton
momentum $\vec{q}$ and the $z$ axis is shown in
Fig.~\ref{fig:LTexcitons}.

Then, the corresponding solutions of Eq.~(\ref{Psystem=}) are
$\vec{P}\propto\boldsymbol{\mu}_\nu^T$ and
$\vec{P}\propto\boldsymbol{\mu}_\nu^L$, and the shifted (complex)
excitonic energies are approximately given by
\begin{align}\label{GoldenRule=}
\tilde{E}_\nu^{T,L}&=E_\nu-(\mu_{\nu{i}}^{T,L})^*
\bar{D}_{ij}(\vec{q},E_\nu)\,\mu_{\nu{j}}^{T,L}\equiv \nonumber \\
&\equiv E_\nu+\Omega_\nu-\frac{i\Gamma_\nu}2.
\end{align}
Here, $-(\mu_{\nu{i}}^{T,L})^*
\bar{D}_{ij}(\vec{q},E_\nu)\,\mu_{\nu{j}}^{T,L}$ is nothing
but the radiative self-energy of the exciton, in which we
explicitly separate the radiative shift $\Omega_\nu$ and the
decay rate $\Gamma_\nu$.
This procedure is equivalent to the quantum-mechanical calculation
of the energy shift by perturbation theory to the first order in
the Coulomb exchange and to the second order in the exciton-photon
coupling, and to the calculation of the radiative decay rate by
the Fermi Golden Rule.

If the spin mixing is due to the Zeeman effect in an in-plane
magnetic field~$\vec{B}_\|$ instead of the Rashba effect, the
$s$ and $p$ polarizations do not separate, as $\vec{q}$
and $\vec{B}_\|$ define two different directions in the plane.
We discuss this situation in Appendix~\ref{app:Zeeman}.

\subsection{Momentum-dependent radiative shifts and decay rates}
\label{ssec:Gammaq}

For the transverse bright and dark $A$ excitons, the radiative
shift and decay rate are straightforwardly evaluated as
\begin{equation}\label{GammaTA=}
\Omega_{TA}^\mathrm{b,d}(\vec{q})
\,-\frac{i}2\,\Gamma_{TA}^\mathrm{b,d}(\vec{q})
\!=\!\left\{\!\!\!\begin{array}{c} 1-w_\vec{q} \\ w_\vec{q} \end{array}\!\!\!\right\}
\frac{8\pi}i\,\frac{(e^2/c)v^2\Phi^2(0)}{cq_{1z}^A+cq_{2z}^A},
\end{equation}
where the upper/lower line in the braces with
$w_\vec{q}\equiv|\lambda_\mathrm{ex}|^2q^2/(4\Delta_c^2)$
refers to the
bright/dark exciton, and the expression to the right of the
braces is the decay rate of the transverse bright $A$~exciton
in the absence of spin flip
\cite{Glazov2014,Gartstein2015,Palummo2015,Wang2016}.
In the square roots,
$q_{1z,2z}^A=\sqrt{\vep_{1,2}E_A^2/c^2-q^2}$,
we neglected the small difference between
$E_A^\mathrm{b}$ and $E_A^\mathrm{d}$ as well as their
dependence on~$\vec{q}$, since
we are interested only in a
narrow region $q<\max\{\sqrt{\vep_1},\sqrt{\vep_2}\}E_A/c$
where at least one of $q_{1z}$, $q_{2z}$ is
real~\cite{footnote_3,footnote_4}.
For the longitudinal $A$ excitons we obtain more bulky
expressions,
\begin{align}
\Omega_{LA}^\mathrm{b}(\vec{q})-&\frac{i}2\,\Gamma_{LA}^\mathrm{b}(\vec{q})
=-\frac{8\pi i\Phi^2(0)}{q_{1z}^A/\vep_1+q_{2z}^A/\vep_2}
\frac{e^2v^2}{E_A^2}
\frac{q_{1z}^Aq_{2z}^A}{\vep_1\vep_2}-{}\nonumber\\
&-\frac{8\pi q^2\Phi^2(0)}{q_{1z}^A/\vep_1+q_{2z}^A/\vep_2}
\frac{ev}{E_A}\frac{\lambda_\mathrm{ex}}{2\Delta_\mathrm{c}} \times \nonumber \\
&\times\left[\left(\frac{q_{1z}^A}{\vep_1}-\frac{q_{2z}^A}{\vep_2}\right)d_z
-i\,\frac{ev}{E_A}\frac{\lambda_\mathrm{ex}}{2\Delta_\mathrm{c}}
\frac{q_{1z}^Aq_{2z}^A}{\vep_1\vep_2}\right],\label{GammaLAb=}\\
\Omega_{LA}^\mathrm{d}(\vec{q})-&\frac{i}2\Gamma_{LA}^\mathrm{d}(\vec{q})
=-\frac{8\pi iq^2\Phi^2(0)}{q_{1z}^A/\vep_1+q_{2z}^A/\vep_2}\times \nonumber \\ &\times
\left[\frac{e^2v^2}{E_A^2}
\frac{\lambda_\mathrm{ex}^2}{4\Delta_\mathrm{c}^2}
\frac{q_{1z}^Aq_{2z}^A}{\vep_1\vep_2}+d_z^2\right]+\nonumber\\
&+\frac{8\pi q^2\Phi^2(0)}{q_{1z}^A/\vep_1+q_{2z}^A/\vep_2}
\frac{ev}{E_A}\frac{\lambda_\mathrm{ex}}{2\Delta_\mathrm{c}}
\left(\frac{q_{1z}^A}{\vep_1}-\frac{q_{2z}^A}{\vep_2}\right)d_z + \nonumber \\
&+4\pi\kappa_0d_z^2\Phi^2(0).\label{GammaLAd=}
\end{align}
whose qualitative features are similar to the transverse case:
the decay rates are nonzero only in the small-momentum region
$q<\max\{\sqrt{\vep_1},\sqrt{\vep_2}\}E_A/c$, the bright exciton
radiative rate is dominated by the first line of
Eq.~(\ref{GammaLAb=}) to which the Rashba term gives a small
correction.
However, an important difference from the transverse case is
that $\Gamma_{LA}^\mathrm{d}$ does not vanish for
$\lambda_\mathrm{ex}=0$. The longitudinal dark $A$ exciton can
decay even in the absence of the Rashba coupling, due to~$d_z$.
Moreover, this latter mechanism is by far the dominant one,
as the ratio between the two terms in the square brackets
of~(\ref{GammaLAd=}) is $\sim{10}^{-5}$ for
a quite high electric field of $\mathcal{E}_z=0.1\:\mbox{V/\AA}$.
Another important difference from the transverse case is that
far in the non-radiative zone, $q\gg\sqrt{\vep_{1,2}}E_A/c$,
the longitudinal exciton energies
$\Omega_{LA}^\mathrm{b,d}(\vec{q})\propto{q}$. This is the
effect of the exchange interaction, which was discussed
in~\cite{Yu2014} (in that work, the intravalley part of the
exchange interaction was not taken into account, which was
later corrected in~\cite{Yu2015,Gartstein2015}).

The last term in Eq.~(\ref{GammaLAd=}) containing the unknown
parameter~$\kappa_0$ is purely real and does not contribute to
the radiative decay rate. However, it produces an energy shift,
\begin{equation}
\Xi_0=4\pi\kappa_0d_z^2\Phi^2(0),
\end{equation}
of the longitudinal
excitons, which thus lifts the valley degeneracy even at
$\vec{q}=0$ and in the absence of Rashba or Zeeman effects
\cite{footnote_5}.
In the macroscopic framework, used here, $\Xi_0$ represents
the interaction energy of the exciton polarization in the
$z$~direction with its own electric field, which is singular
in the limit of an infinitely thin layer.
By the order of magnitude, $\kappa_0\sim\vep/d$, where
$d$~is the monolayer thickness and $\vep\approx{7}$ is the
effective background dielectric constant in the perpendicular
direction at optical frequencies (see Appendix~\ref{app:Maxwell}).
Taking $d=3\:\mbox{\AA}$, $\vep=3$,
$d_z/e=0.08\:\mbox{\AA}$, and
$\Phi(0)\sim{0}.1\:\mbox{\AA}^{-1}$, we obtain an estimate
$\Xi_0\sim{10}\:\mbox{meV}$.
This value agrees by the order of magnitude with the dark
exciton energy shifts due to the short-range exchange
interaction, which were calculated using a microscopic
\textit{ab initio} approach in \cite{Echeverry2016}.
Indeed, both are supposed to have the same origin, as the
Coulomb part of $\bar{D}_{ij}(\vec{q},\omega)$ represents
the exchange field. However, no lifting of valley degeneracy
was mentioned in \cite{Echeverry2016}.
Note that the presence of the positive shift $\Xi_0$ for the
longitudinal excitons has an important consequence for the
luminescence, as the lowest-energy states are the transverse
excitons whose decay rate is only due to Rashba or Zeeman
effect.

\begin{figure}[]
\includegraphics[width=8.5cm]{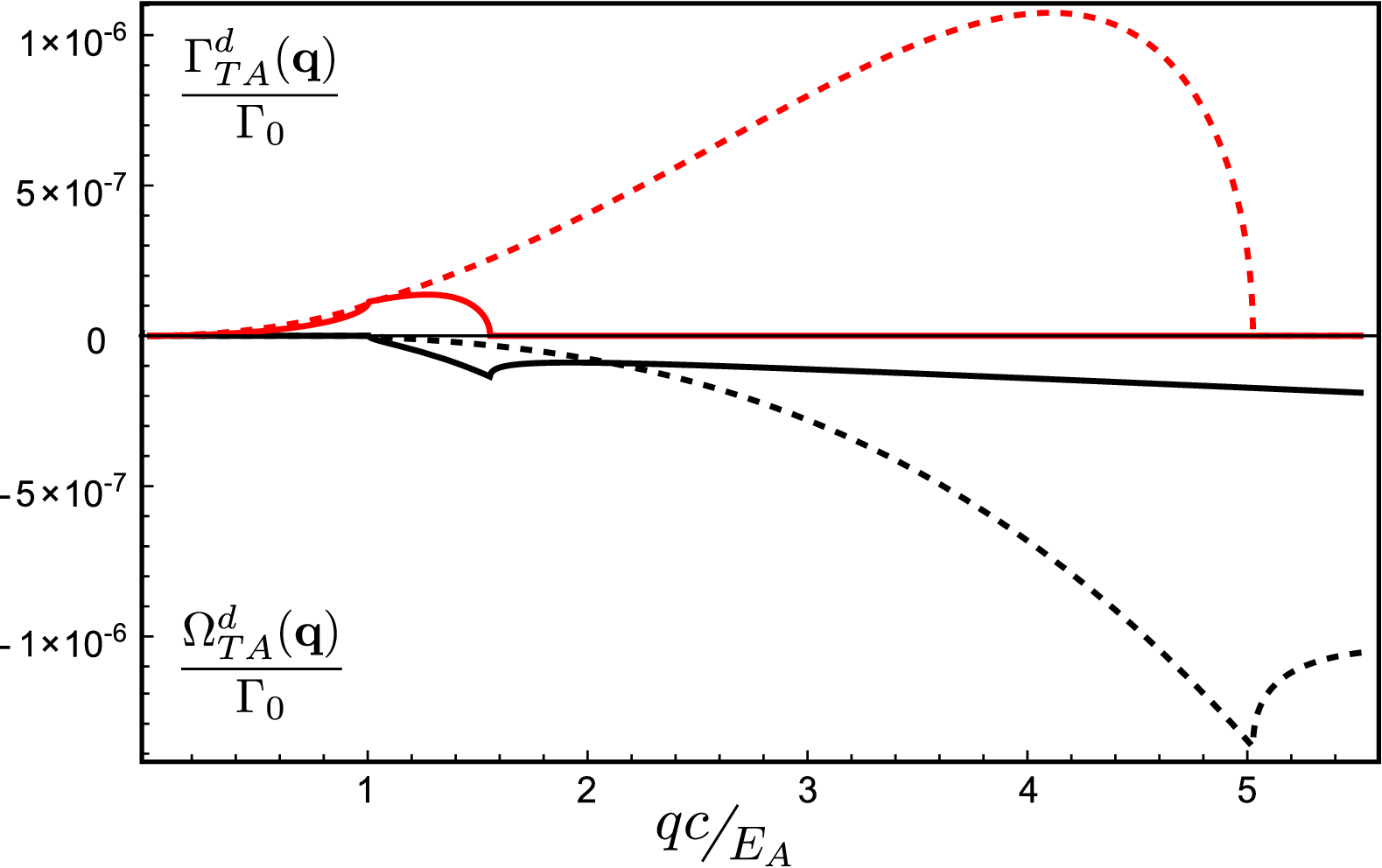}
\caption{
The normalized radiative energy shifts
$\Omega^d_{TA}(\vec{q})/\Gamma_0$
(black curves, negative $y$~axis)
and decay rates $\Gamma^d_{TA}(\vec{q})/\Gamma_0$
(red curves, positive $y$~axis) of transverse excitons for
$\varepsilon_1=1$, $\varepsilon=2.4$ (solid curves) and
$\varepsilon=25$ (dashed curves),
as a function of the dimensionless exciton momentum $cq/E_A$.}
\label{fig:TA}
\end{figure}

\begin{figure}[]
\includegraphics[width=8.5cm]{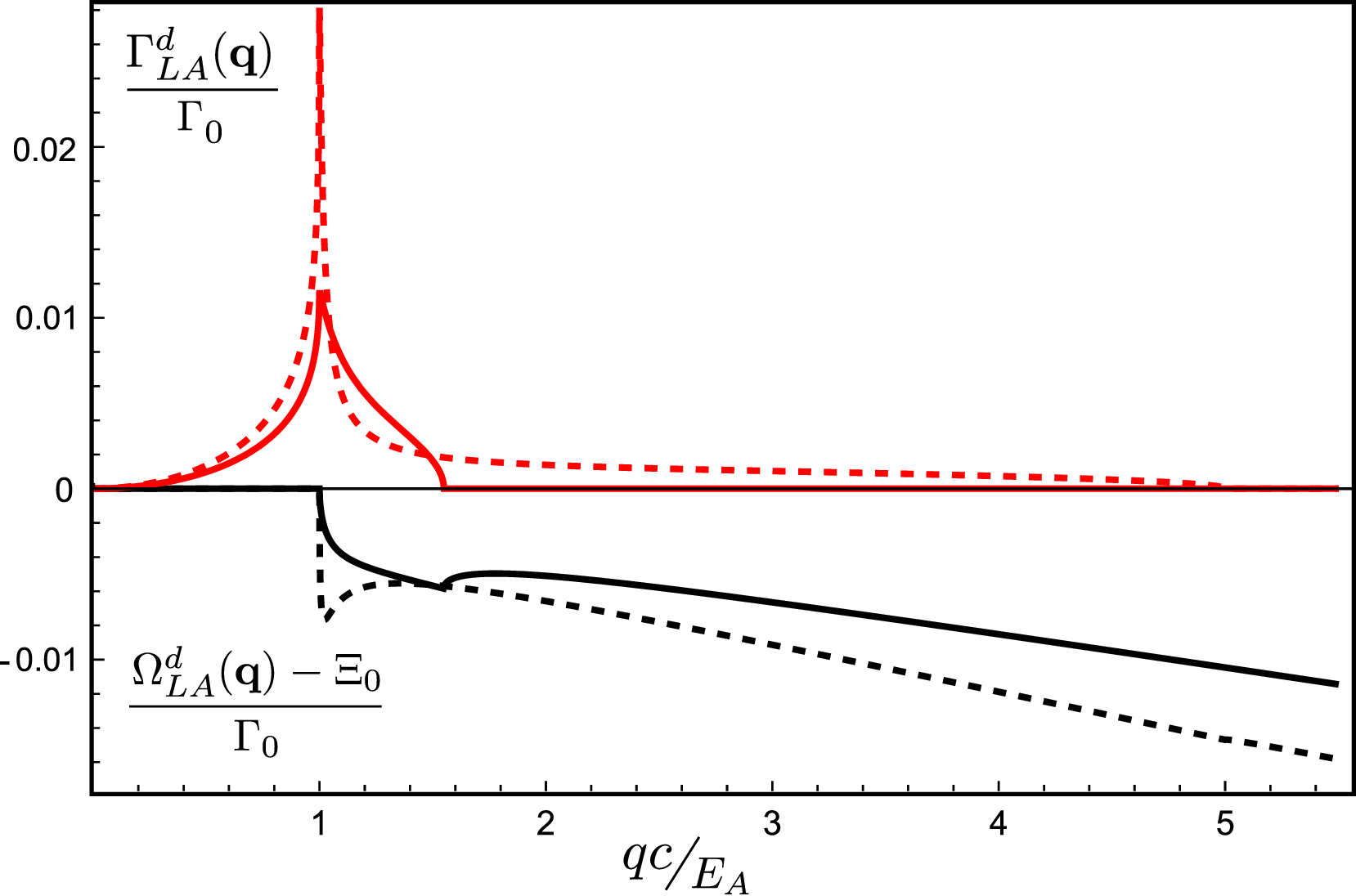}
\caption{
The normalized radiative energy shifts (excluding the short-range
exchange spliting~$\Xi_0$)
$[\Omega^d_{LA}(\vec{q})-\Xi_0]/\Gamma_0$
(black curves, negative $y$~axis)
and decay rates $\Gamma^d_{LA}(\vec{q})/\Gamma_0$
(red curves, positive $y$~axis) of longitudinal excitons
for $\varepsilon_1=1$, $\varepsilon=2.4$ (solid curves) and
$\varepsilon=25$ (dashed curves),
as a function of the dimensionless exciton momentum $cq/E_A$.
}
\label{fig:LA}
\end{figure}

To illustrate the $q$ dependence of the radiative energy
shifts and the decay rates, we plot them in
Figs.~\ref{fig:TA} and~\ref{fig:LA} for the transverse and
longitudinal dark $A$ excitons, for $\vep_1=1$ and two values
of the substrate dielectric constant, one typical for
glass-like substrates, $\vep_2=2.4$, the other one
corresponding to a highly dielectric substrate, such as
AlGaSb with $\vep_2=25$ at optical frequencies~\cite{Ferrini1998}.
It is convenient to normalize the energy shifts and decay
rates by the bright exciton decay rate at $\vec{q}=0$,
$\Gamma_{TA}^\mathrm{b}(0)=\Gamma_{LA}^\mathrm{b}(0)\equiv\Gamma_0$,
for a TMDC monolayer suspended in vacuum,
\begin{equation}\label{Gamma0=}
\Gamma_0=8\pi\,\frac{e^2}c\,\frac{v^2\Phi^2(0)}{E_A}.
\end{equation}
For $v=2.6\:\mbox{eV}\cdot\mbox{\AA}$,
$E_A=1.7\:\mbox{eV}$, $\Phi(0)=0.1\:\mbox{\AA}^{-1}$,
this estimate gives
$\Gamma_0\approx{7}\:\mbox{meV}\approx(100\:\mbox{fs})^{-1}$,
not very much different from the results of more precise
calculations involving the full microscopic treatment of the
exciton wave function~\cite{Palummo2015,Wang2016}, which
give $1/\Gamma_0\approx{2}00\:\mbox{fs}$ for several TMDC
materials.
In the calculation presented in Figs.~\ref{fig:TA},\ref{fig:LA},
we used the values $E_A=1.7\:\mbox{eV}$,
the experimentally determined splitting
$2\Delta_\mathrm{c}=-30\:\mbox{meV}$~\cite{Zhang2015},
$d_z/e=0.08\:\mbox{\AA}$ as extracted from
\cite{Kormanyos2014},
and the Rashba coupling
constant $\lambda_\mathrm{ex}=90\:\mbox{meV}\cdot\mbox{\AA}$
per each V/{\AA} of the static electric
field~$\mathcal{E}_z$~\cite{Kormanyos2014}.
The latter was taken to be 0.1~V/{\AA} for the glass-like
substrate, giving
$\lambda_\mathrm{ex}=9\:\mbox{meV}\cdot\mbox{\AA}$,
and 0.2~eV/{\AA} for the highly dielectric substrate
($\lambda_\mathrm{ex}=18\:\mbox{meV}\cdot\mbox{\AA}$),
as estimated in Appendix~\ref{app:substrate} (note that
changing the electric field amounts to a simple rescaling
of the $y$~axis $\propto\mathcal{E}_z^2$ in Fig.~\ref{fig:TA},
while Fig.~\ref{fig:LA} is insensitive to~$\mathcal{E}_z$, as
discussed above).

\subsection{Integrated decay rates}
\label{ssec:integrated}

So far, we calculated the decay rates $\Gamma_\nu(\vec{q})$
for the excitonic states with a given momentum~$\vec{q}$
and these rates depend quite strongly on~$q$. It is interesting
to compare the overall efficiency of different decay mechanisms,
studied here, with respect to the radiative decay of bright
excitons. To do this, we integrate the obtained decay rates
over~$\vec{q}$.
For bright excitons, integration of Eqs.~(\ref{GammaTA=}) and
(\ref{GammaLAb=}) gives
\begin{equation}
\int\frac{d^2\vec{q}}{(2\pi)^2}\,
\frac{\Gamma^\mathrm{b}_{TA}(\vec{q})
+\Gamma^\mathrm{b}_{LA}(\vec{q})}{\Gamma_0}=
\frac{E_A^2}{2\pi{c^2}}\,2\mathcal{F}_\mathrm{b}(\vep_1,\vep_2),
\end{equation}
where $\mathcal{F}_\mathrm{b}(\vep_1,\vep_2)$ is a function of
the two dielectric constants, whose explicit form is given in
Appendix~\ref{app:integral}.
For dark exciton decay due to the Rashba spin-orbit coupling,
we integrate Eqs.~(\ref{GammaTA=}) and
(\ref{GammaLAd=}) and obtain
\begin{align}
\int\frac{d^2\vec{q}}{(2\pi)^2}\,
\frac{\Gamma^\mathrm{d}_{LA}(\vec{q})}{\Gamma_0}=
\frac{E_A^2}{2\pi{c}^2}\,
\eta_{d_z}\mathcal{F}_\mathrm{d}^L(\vep_1,\vep_2),\\
\int\frac{d^2\vec{q}}{(2\pi)^2}\,
\frac{\Gamma^\mathrm{d}_{TA}(\vec{q})}{\Gamma_0}=
\frac{E_A^2}{2\pi{c}^2}\,\eta_\mathrm{R}
\mathcal{F}^\mathrm{d}_T(\vep_1,\vep_2),
\end{align}
while in the presence of the in-plane Zeeman field we
integrate Eqs.~(\ref{GammaBpar=}), (\ref{GammaBperp=})
of Appendix~\ref{app:Zeeman}:
\begin{align}
&\int\frac{d^2\vec{q}}{(2\pi)^2}\,
\frac{\Gamma^\mathrm{d}_\|(\vec{q})}{\Gamma_0}=
\frac{E_A^2}{2\pi{c}^2}
\left[\eta_{d_z}\mathcal{F}_\mathrm{d}^L(\vep_1,\vep_2)
+\eta_\mathrm{Z}\,\mathcal{F}_\mathrm{b}(\vep_1,\vep_2)\right],\\
&\int\frac{d^2\vec{q}}{(2\pi)^2}\,
\frac{\Gamma^\mathrm{d}_\perp(\vec{q})}{\Gamma_0}=
\frac{E_A^2}{2\pi{c}^2}\,\eta_\mathrm{Z}\,
\mathcal{F}_\mathrm{b}(\vep_1,\vep_2).
\end{align}
The explicit form of the functions
$\mathcal{F}^\mathrm{d}_T(\vep_1,\vep_2)$,
$\mathcal{F}^\mathrm{d}_L(\vep_1,\vep_2)$,
$\mathcal{F}^\mathrm{b}(\vep_1,\vep_2)$
is given in Appendix~\ref{app:integral}.
For $\vep_1=1$, $\vep_2=2.4$, they are equal to
1.5, 1.6, and 1.0, respectively.
The factors $\eta_{d_z}$, $\eta_\mathrm{R}$ and
$\eta_\mathrm{Z}$ are defined as
\begin{align}
\label{etadz=}
&\eta_{d_z}=\frac{d_z^2}{e^2}\left(\frac{E_A}v\right)^2,\\
\label{etaR=}
&\eta_\mathrm{R}=\frac{\lambda_\mathrm{ex}^2}%
{4\Delta_\mathrm{c}^2}\left(\frac{E_A}c\right)^2,\\
&\eta_\mathrm{Z}=
\left(\frac{g_\|\mu_BB_\|}{4\Delta_\mathrm{c}}\right)^2.
\label{etaZ=}
\end{align}
Their values are estimated to be
$\eta_{d_z}\approx{3}\times{10}^{-3}$,
$\eta_\mathrm{R}\approx{0}.7\times{10}^{-7}$ for the
perpendicular electric field
$\mathcal{E}_z=0.1\:\mbox{V/\AA}$,
and $\eta_\mathrm{Z}\approx{4}\times{10}^{-3}$
for the in-plane magnetic field $B_\|=30\:\mbox{T}$
and $g_\|=-2$.
These estimates represent the main result of the present
paper. Indeed, since all
$\mathcal{F}^\mathrm{b,d}(\vep_1,\vep_2)\sim{1}$,
the relative importance of different decay mechanisms is
mainly determined by the factors $\eta_\mathrm{R}$,
$\eta_{d_z}$, and $\eta_\mathrm{Z}$.

The momentum-integrated decay rates, calculated above,
besides giving a simple estimate of relative importance
of each decay mechanism, also determine the total
oscillator strgenth of each exciton species, which plays
a role in different physical situations.
One such situation is a disordered sample, where exciton
center-of-mass motion is no longer free, so its wave
function is not a plane wave $e^{i\vec{q}\vec{r}}$. If,
instead, the exciton state has some complicated
center-of-mass wave function $\Psi(\vec{r})$, its decay
rate is determined by the weights
$\left|\langle\vec{q}|\Psi\rangle\right|^2$
of different plane waves in this state (here we use the
fact that the exciton-photon coupling is diagonal in the
in-plane momentum~$\vec{q}$):
\begin{equation}
\Gamma[\Psi]=\int\frac{d^2\vec{q}}{(2\pi)^2}\,\Gamma(\vec{q})\,
\left|\langle\vec{q}|\Psi\rangle\right|^2.
\end{equation}
If the characteristic spatial scale of $\Psi(\vec{r})$
(mean free path or localization length) is shorter than
the light wavelength $2\pi{c}/E_A$, one can approximate
$\left|\langle\vec{q}|\Psi\rangle\right|^2$ by a constant
in the narrow radiative region, and then the decay rate is
determined by the momentum integral of~$\Gamma(\vec{q})$.

Another context in which the integral of $\Gamma(\vec{q})$
arises, is the thermal average~\cite{Andreani1991}. Indeed,
if the excitons have the Maxwell-Boltzmann momentum
distribution, $f(\vec{q})\propto{e}^{-q^2/(2m_\mathrm{ex}T)}$,
where $T$~is the temperature and
$m_\mathrm{ex}=m_\mathrm{e}+m_\mathrm{h}$ is the exciton mass,
the typical thermal momenta $q\sim\sqrt{m_\mathrm{ex}T}$
are usually much larger than the radiative momentum $E_A/c$.
Then, the thermal average can be approximated as
\begin{align}
\langle\Gamma_A\rangle
&=\sum_\nu\int\frac{d^2\vec{q}}{(2\pi)^2}\,
\frac{f_\nu(\vec{q})}{n_\mathrm{ex}}\,\Gamma_\nu(\vec{q})\approx\nonumber\\
&\approx\sum_\nu\frac{f_\nu(0)}{n_\mathrm{ex}}
\int\frac{d^2\vec{q}}{(2\pi)^2}\,
\Gamma_\nu(\vec{q}),\label{averageGamma=}
\end{align}
where the summation is over the four species of
$A$~excitons (two bright and two dark), and
$n_\mathrm{ex}$ is their total density. Such average
would represent the global radiative decay rate of the
whole thermal excitonic population, and the relative
contribution of each exciton species would correspond to
the relative intensity of luminescence from each species.
It is in this context that the difference between MoX$_2$
and WX$_2$ becomes important: while the calculation of
$\Gamma_\nu(\vec{q})$ in Sec.~(\ref{ssec:Gammaq}) was
equally applicable to both cases, the lower energy of the
dark excitons in tungsten-based compounds results in a
larger weight of dark excitons in the luminescence. Namely,
the weight of the bright excitons is small by a factor
$e^{-2\Delta_\mathrm{c}/T}$. Similarly, the contribution
of longitudinal dark excitons is small by a factor
$e^{-\Xi_0/T}$ with respect to the transverse contribution.

It should be noted, however, that the equilibrium
average~(\ref{averageGamma=}) is based on the assumption
of the exciton thermalization in the radiative region.
This requires the thermalization rate to be larger
than the radiative rate, in order to quickly supply the
excitons into the narrow radiative region with small
$q\sim{E}_A/c$.
This is not always the case; for example, the full kinetic
treatment of excitons interacting with acoustic phonons in
GaAs quantum wells revealed a relaxation bottleneck at the
border of the radiative region~\cite{Piermarocchi1996}.
To check, whether the excitons in the radiative region
are thermalized, one should compare the radiative
rate~$\Gamma(\vec{q})$ with the acoustic phonon
absorption rate $1/\tau_0$~\cite{Slobodeniuk2016}.
For TMDCs, $1/\tau_0=aT$, where for the constant~$a$
values of a few tens of~$\mu\mbox{eV}/K$ have been reported
\cite{Moody2015,Koirala2016,Dey2016,Selig2016,Jakubczyk2016},
which is much longer than the sub-picosecond radiative rate
for the bright excitons~\cite{Palummo2015,Wang2016} at
temperatures lower than about ${100}\:\mbox{K}$. At lower
temperatures, the bright exciton distribution in the radiative
region is thus expected to be different from the thermal one.
At the same time, for dark excitons whose radiative decay is
at least 100 times slower than for the bright ones (see
Fig.~\ref{fig:LA}), so the thermal approximation should work
reasonably well.

\section{Conclusions and outlook}
\label{sec:conclusions}

In this work we have investigated the radiative decay of
spin-forbidden dark intravalley $A$ excitons in tungsten
dichalcogenide monolayers due to several spin-flip
mechanisms. Our main result is represented by
expressions (\ref{etadz=})--(\ref{etaZ=}) for the average
efficiencies $\eta_\mathrm{R}$, $\eta_z$, and $\eta_\mathrm{Z}$
of different spin-flip mechanisms with respect to the main
radiative decay mechanism of the bright excitons.
As the bright excitons in tungsten-based compounds have
higher energy than the dark ones, the bright exciton
contribution to the photoluminescence
is suppressed at low temperatures by the thermal
activation exponential $e^{-2\Delta_\mathrm{c}/T}$ with
the bright-dark splitting $2\Delta_\mathrm{c}$ of several
tens of meV, of the order of the room temperature.

The intrinsic mechanism of the dark exciton radiative
decay is due to the interband spin-flip dipole moment,
perpendicular to the monolayer plane. It was mentioned
in \cite{Glazov2014,Echeverry2016}, and its
magnitude can be
deduced from the estimates of \cite{Kormanyos2014}.
We found, however, that the very same mechanism lifts
the valley degeneracy of the dark excitons by a Coulomb
local-field effect, producing a splitting $\Xi_0$ which
we could very roughly estimate as about 10~meV. Crucially,
the whole oscillator strength of such interband spin flip
(about $\eta_z\sim{}10^{-2}-10^{-3}$ of the bright exciton
oscillator
strength) is taken by the higher-energy component. Thus,
the spin-forbidden intravalley exciton,
usually referred to as dark, in fact, has two components,
one dark and the other which can be called ``dim''. The
contribution of this dim exciton to the photoluminescence
is exponentially suppressed at low temperatures,
$e^{-\Xi_0/T}$.

For the lowest-energy dark component, which should dominate
the exciton population at low temperatures, we have analyzed
extrinsic spin-flip mechanisms, which transfer some oscillator
strength from the bright exciton to the dark one.
One such mechanism is due to the Rashba spin-orbit coupling
which arises whenever the reflection symmetry in the
monolayer plane is broken, e.~g., by an externally applied
perpendicular static electric field, or by different
dielectric environment above and below the monolayer. This
mechanism was mentioned in \cite{Dery2015}.
Using our own estimate of
$0.1\:\mbox{V/\AA}$ for the effective electric field
produced by the dielectric mismatch and the calculated
value of the Rashba coupling constant from
\cite{Kormanyos2014}, we find that the fraction of the
oscillator strength transferred from the bright exciton to
the dark one is extremely low, about $\eta_\mathrm{R}\sim{}10^{-7}$, mainly due
to small momenta of excitons subject to radiative decay.
Another possible mechanism is the spin flip by the Zeeman
effect from a magnetic field directed along the monolayer
plane.
Then, for a strong but still realistic field of 30~T, a
fraction $\eta_\mathrm{Z}>10^{-3}$ of the bright exciton oscillator strength
can be transferred to the dark one. This suggests a way to
manipulate the radiative properties of dark excitons.

The results of the present work help to identify several
directions for improvement of our understanding of the
excitonic radiative processes in tungsten dichalcogenides.
First, the value of the interband spin-flip dipole moment
is currently known from only one source, the estimate of
\cite{Kormanyos2014}. It would be quite helpful if
more information were available, either theoretically or
experimentally. e.~g., by studying the effect of the Rashba
spin-orbit coupling on the carrier transport, as the Rashba
coupling has the same origin.
Second, a reliable microscopic calculation of the dark-dim
exciton splitting~$\Xi_0$ would help to determine how
dramatic is the low-temperature suppression of the dim
exciton population.
Finally, the role of collisions with defects, charge
carriers, or phonons in the radiative decay also needs
clarification.

\section{Aknowledgements}

We are grateful to M.~Potemski for stimulating discussions.
A. O. S. acknowledges financial support from
the EC Graphene Flagship project (No. 604391).

\appendix

\section{Interband spin-flip dipole matrix element}
\label{app:dipole_z_term}


To fix the form of~$d_z$ in Eq.~(\ref{HRashba=}), let us start from the Bloch states of TMDC monolayer obtained without taking into account the spin-orbit coupling. The construction involves the top valence band ($\mathrm{v}$), the lowest conduction band ($\mathrm{c}$), and a third band ($\mathrm{z}$), which we take to be the next highest conduction band~\cite{Kormanyos2014}. The coordinate wave functions of the corresponding states at $\pm\vec{K}$ points can be written as $u_{\mathrm{v},\pm\vec{K}}(\vec{r})\,e^{\pm{i}\vec{K}\vec{r}}$, $u_{\mathrm{c},\pm\vec{K}}(\vec{r})\,e^{\pm{i}\vec{K}\vec{r}}$,
$u_{\mathrm{z},\pm\vec{K}}(\vec{r})\,e^{\pm{i}\vec{K}\vec{r}}$. With respect to the reflection $z\to-z$ in the crystal plane, the functions $u_{\mathrm{v},\pm\vec{K}}$ and $u_{\mathrm{c},\pm\vec{K}}$ are even, while $u_{\mathrm{z},\pm\vec{K}}$ is odd.
Time reversal symmetry of the Hamiltonian without the spin-orbit coupling imposes
\begin{eqnarray}
&u_{\mathrm{z},-\vec{K}}=e^{i\varphi_\mathrm{z}}u_{\mathrm{z},\vec{K}}^*,\label{eq:T1}\\
&u_{\mathrm{c},-\vec{K}}=e^{i\varphi_\mathrm{c}}u_{\mathrm{c},\vec{K}}^*,\\
&u_{\mathrm{v},-\vec{K}}=e^{i\varphi_\mathrm{v}}u_{\mathrm{v},\vec{K}}^*,\label{eq:T3}
\end{eqnarray}
where $u^*$ denotes the complex conjugate of~$u$ and $\varphi_{\mathrm{v},\mathrm{c},\mathrm{z}}$ are some phases which we can choose freely. At the same time, reflection symmetry $x\to-x$ requires
\begin{eqnarray}
&u_{\mathrm{z},-\vec{K}}=e^{i\phi_\mathrm{z}}\bar{u}_{\mathrm{z},\vec{K}},\label{eq:X1}\\
&u_{\mathrm{c},-\vec{K}}=e^{i\phi_\mathrm{c}}\bar{u}_{\mathrm{c},\vec{K}},\\
&u_{\mathrm{v},-\vec{K}}=e^{i\phi_\mathrm{v}}\bar{u}_{\mathrm{v},\vec{K}},\label{eq:X3}
\end{eqnarray}
where we denote $\bar{u}(x,y,z)=u(-x,y,z)$, and $\phi_{\mathrm{v},\mathrm{c},\mathrm{z}}$ are some other phases. One can choose either the $\varphi$~phases or the $\phi$ phases, but as soon as one set is fixed, the other one is fixed too.
The requirement for the $\vec{k}\cdot\vec{p}$ perturbation theory to give a Dirac-like coupling between the valence and the conduction band with a real velocity ($v$~in Eq.~\ref{Htauk=}) translates into the following condition:
\begin{equation}\label{eq:kp}
\langle{u}_{\mathrm{c},\pm\vec{K}}|(-i\hbar\boldsymbol\nabla/m)|
u_{\mathrm{v},\pm\vec{K}}\rangle=v(\pm\vec{e}_x-i\vec{e}_y),
\end{equation}
where the bracket notations stand for the integration in the coordinate space:
\begin{equation}
\langle{u}_1|\hat{O}|u_2\rangle\equiv
\int{u}_1^*(\vec{r})\,\hat{O}\,u_2(\vec{r})\,d^3\vec{r}.
\end{equation}
Condition (\ref{eq:kp}) fixes $\varphi_\mathrm{c}=\varphi_\mathrm{v}$, $\phi_\mathrm{c}=\phi_\mathrm{c}$.

The spin structure is described by the two spinors
\[
\chi_\uparrow=\left(\begin{array}{c} 1 \\ 0 \end{array}\right),\quad
\chi_\downarrow=\left(\begin{array}{c} 0 \\ 1 \end{array}\right),
\]
so the full wave functions in each band become $u_{\pm\vec{K}}(\vec{r})\,e^{\pm{i}\vec{K}\vec{r}}\chi_\uparrow$, $u_{\pm\vec{K}}(\vec{r})\,e^{\pm{i}\vec{K}\vec{r}}\chi_\downarrow$. The Zeeman term is $g\mu_B\vec{B}\cdot\boldsymbol\sigma/2$, where the Pauli matrices act on the spinors. The spin-orbit coupling introduces the spin splitting terms $\pm\Delta_{\mathrm{v},\mathrm{c}}\sigma_z$ for the valence/conduction band at the $\pm\vec{K}$ point. When both $\Delta_{\mathrm{v},\mathrm{c}}>0$, the top valence band wave function at the $\vec{K}$ point is $u_{\mathrm{v},\vec{K}}e^{i\vec{K}\vec{r}}\chi_\uparrow$, and the bottom conduction band wave function is $u_{\mathrm{c},\vec{K}}e^{i\vec{K}\vec{r}}\chi_\downarrow$, so the optical transition requires a spin flip.

The spin-flip transition dipole moment $d_z$ is constructed from the second-order perturbation theory involving a virtual transition to the band ($\mathrm{z}$), once with the $z$ operator, and once with the spin-orbit coupling $\propto \hat{L}_+\sigma_-+\hat{L}_-\sigma_+$, where $\hat{L}_\pm=\hat{L}_x\pm{i}\hat{L}_y$. Denoting the corresponding matrix element in the respective valleys by $d_{\pm\vec{K}}$, we can write (up to constant factors)
\begin{eqnarray}
d_{\pm\vec{K}}\sim
\langle{u}_{\mathrm{v},\pm\vec{K}}|z|u_{\mathrm{z},\pm\vec{K}}\rangle
\langle{u}_{\mathrm{z},\pm\vec{K}}|\hat{L}_\mp|u_{\mathrm{c},\pm\vec{K}}\rangle+{}\nonumber\\ \qquad\quad
{}+\langle{u}_{\mathrm{v},\pm\vec{K}}|\hat{L}_\mp|u_{\mathrm{z},\pm\vec{K}}\rangle
\langle{u}_{\mathrm{z},\pm\vec{K}}|z|u_{\mathrm{c},\pm\vec{K}}\rangle,
\end{eqnarray}
where only the coordinate parts of the wave functions enter (the spin part of the expression has been evaluated explicitly).

If we now substitute $u_{-\vec{K}}$ for all bands from equations~(\ref{eq:T1})--(\ref{eq:T3}) with  $\varphi_\mathrm{c}=\varphi_\mathrm{v}$ following from~(\ref{eq:kp}) and use $\hat{L}_-^*=-\hat{L}_+$ in the coordinate representation, we obtain $d_\vec{-K}=-d_\vec{K}^*$. 
At the same time, using equations~(\ref{eq:X1})--(\ref{eq:X3}) and noting that $z$~remains invariant upon reflection $x\to-x$ while $\hat{L}_+\to\hat{L}_-$, we obtain $d_\vec{-K}=d_\vec{K}$. As a result, $d_\vec{-K}=d_\vec{K}=id_z$ is purely imaginary, as in~\cite{Ochoa2013}.

\section{Block-diagonal form of the Rashba coupling}
\label{app:diag_hamiltonian}

The $4\times4$ matrix
$\mathcal{H}_\tau(\vec{k})-\mathcal{E}_z\mathcal{D}_\tau$,
can be transformed to the block-diagonal intra-band form
by applying a unitary transformation
$e^{-\mathcal{S}_\tau(\vec{k})}$
(here $\mathcal{E}_z$ is only the static electric field,
while the same term with the optical field remains inter-band):
\begin{equation}
\tilde{\mathcal{H}}_\tau(\vec{k})+\mathcal{H}^\mathrm{R}(\vec{k})
=e^{-\mathcal{S}_\tau(\vec{k})}
\left[\mathcal{H}_\tau(\vec{k})-\mathcal{E}_z\mathcal{D}_\tau\right]
e^{\mathcal{S}_\tau(\vec{k})}
\end{equation}
If the matrix $\mathcal{S}_\tau(\vec{k})$ is chosen in the form
\begin{align}
&\mathcal{S}_{+1}(\vec{k})=\frac{1}{E_\mathrm{g}}\left[\begin{array}{cccc}
0 & 0 & -vk_- & 0 \\
0 & 0 & -i\mathcal{E}_zd_z & -vk_- \\
vk_+ & -i\mathcal{E}_zd_z & 0 & 0 \\
0 & vk_+ & 0 & 0 \end{array}\right],\\
&\mathcal{S}_{-1}(\vec{k})=\frac{1}{E_\mathrm{g}}\left[\begin{array}{cccc}
0 & 0 & vk_+ & -i\mathcal{E}_zd_z \\
0 & 0 & 0 & vk_+ \\ -vk_- & 0 & 0 & 0 \\
-i\mathcal{E}_zd_z & -vk_- & 0 & 0 \end{array}\right],
\end{align}
where $k_\pm=k_x\pm i k_y$, then, to the second order in
$\mathcal{E}_z|d_z|/E_\mathrm{g}\ll{1}$, $vk/E_\mathrm{g}\ll{1}$,
we obtain a block-diagonal Hamiltonian. The first term represents
the kinetic energy of the electrons and holes,
\begin{equation}
\tilde{\mathcal{H}}_\tau(\vec{k})=\left[\begin{array}{cc}
E_\mathrm{g}+\tau\Delta_\mathrm{c}\sigma_z+\frac{k^2}{2m_\mathrm{e}} & 0 \\
0 & \tau\Delta_\mathrm{v}\sigma_z-\frac{k^2}{2m_\mathrm{h}}\end{array}\right],
\end{equation}
while the second one corresponds to the Rashba coupling,
\begin{equation}
\mathcal{H}^\mathrm{R}(\vec{k})=\frac{v\mathcal{E}_z}{E_\mathrm{g}}d_z
\left[\begin{array}{cccc}
0 & -ik_- & 0 & 0 \\ ik_+ & 0 & 0 & 0 \\
0 & 0  & 0 & ik_- \\ 0 & 0 & -ik_+ & 0 \end{array}\right].
\end{equation}
Here we also assumed $|\Delta_\mathrm{c,v}|\ll{E}_\mathrm{g}$.
TMDC materials can have comparable values of the spin splitting in
the valence band $\Delta_\mathrm{v}$ and the gap $E_\mathrm{g}$;
then the above calculation can be repeated without assuming
$|\Delta_\mathrm{c,v}|\ll{E}_\mathrm{g}$, which would produce
different Rashba coupling in the conduction and the valence band.
For tungsten compounds $\Delta_\mathrm{v}/E_\mathrm{g}\sim0.1$,
so we neglect this difference.
We also neglect the effect of the unitary transformation
$e^{-\mathcal{S}_\tau(\vec{q})}$ on the Coulomb interaction which forms
the excitonic states.

\section{Estimate of the effective substrate-induced electric field}
\label{app:substrate}

An external electric field $\mathcal{E}_z$ and interaction with
a substrate via van der Waals forces have the same qualitative
effect of breaking the reflection symmetry in the $z$~direction
and deforming the electronic wave functions in the TMDC monolayer.
Still, there is no reason for the deformation to be quantitatively
similar in the two cases. Thus, characterization of the substrate
effect by a single parameter, an effective~$\mathcal{E}_z$, is quite
a rough approximation, which gives only an order-of-magnitude
estimate of the effect.
Furthermore, we adopt a macroscopic description of the substrate,
characterizing it by the dielectric constant. As the substrate is
only a few angstroms away from the monolayer center, such macroscopic
description is also valid only qualitatively.

We model the TMDC monolayer as a slab of thickness $d$ and dielectric
constant~$\varepsilon$, sandwiched between two media with dielectric
constants $\varepsilon_1$ at $z>d/2$ and $\varepsilon_2$ at $z<-d/2$.
Let us write the energy as a functional of the microscopic
three-dimensional electronic density $\rho(\vec{R})$,
$\vec{R}\equiv(\vec{r},z)$ where the
substrate effect is included as a Hartree-like term:
\begin{align}
E[\rho(\vec{R})]={}&{}E_0[\rho(\vec{R})]-{}\nonumber\\
{}&{}-\int\left[V(\vec{R},\vec{R}')-V_1(\vec{R},\vec{R}')\right]
\times\nonumber\\ {}&\qquad{}\times
\rho(\vec{R})\,\rho_\mathrm{i}(\vec{R}')\,d^3\vec{R}\,d^3\vec{R}'+\nonumber\\
{}&{}+\frac{1}{2}\int
\left[V(\vec{R},\vec{R}')-V_1(\vec{R},\vec{R}')\right]
\times\nonumber\\ {}&\qquad{}\times
\rho(\vec{R})\,\rho(\vec{R}')\,d^3\vec{R}\,d^3\vec{R}'.
\label{Hartree=}
\end{align}
Here $E_0[\rho(\vec{R})]$ is the functional for a TMDC monolayer
suspended in vacuum, $\rho_\mathrm{i}(\vec{R}')$ is the ionic density
(a sum of $\delta$~functions at the ion positions),
$V(\vec{R},\vec{R}')$ is the interaction potential, represented by
the Green's function of the Poisson equation in the full dielectric
structure multiplied by~$e^2$,
$V_1(\vec{R},\vec{R}')$ is the same for $\vep_1=\vep_2=1$.
The second and the third terms in Eq.~(\ref{Hartree=}) describe
interaction of the TMDC electrons with the polarization charges in the
surrounding media. The deformed electronic wave functions can be found
by varying the functional~(\ref{Hartree=}). The whole description is
analogous to the macroscopic description of polarons in ionic crystals
\cite{Pekar1946}; indeed, deformation of electronic wave functions
in the TMDC by interaction with a dielectric substrate can be viewed
as a polaronic effect.

In the planar geometry considered here, evaluation of the integrals in
Eq.~(\ref{Hartree=}) reduces to a summation over image charges. When
all three dielectric constants $\vep,\vep_1,\vep_2$ are different,
each charge produces an infinite number of images:
\begin{align}
&V(\vec{R},\vec{R}')=\!\!\!\!\sum_{n=-\infty}^\infty\frac{\zeta_ne^2/\vep}%
{l_n},\\
&l_n=\sqrt{|\vec{r}-\vec{r}'|^2+[z-(-1)^nz'-nd]^2},\\
&\zeta_0=1,\,
\zeta_{\pm{1}}=\frac{\vep-\vep_{1,2}}{\vep+\vep_{1,2}},\,
\zeta_{\pm{2}}=\zeta_{\pm{1}}\,
\frac{\vep-\vep_{2,1}}{\vep+\vep_{2,1}},\,\ldots
\end{align}
For $\vep_1=\vep_2=1\ll\vep$, the decay of $\zeta_n$ with $|n|$
is quite slow (which is a manifestation of the confinement of
the electric field lines to the interior of the dielectric),
so many images contribute to the Hartree terms in
Eq.~(\ref{Hartree=}). However, if we start with $\rho(\vec{R})$,
symmetric with respect to $z\to-z$, the image charge distribution
due to~$V_1$ is also symmetric and does not produce any net
electric field in the $z$~direction. As concerns the $V$~term,
if at least one of $\vep_1$, $\vep_2$ is not small compared
to~$\vep$, which we assume to be the case, $\zeta_n$~decays
exponentially with a decrement $\sim{1}$. Thus, for an
order-of-magnitude estimate, we can restrict ourselves
to~$\zeta_{\pm{1}}$.

The electronic states we are interested in (the bottom of the
conduction band and the top of the valence band at $\pm\vec{K}$
points) are known to originate mainly from $d$~orbitals of the
metal atoms. Assuming these orbitals to be concentrated around
$z=0$, we can expand the potentials in $z/d$ as a formal small
parameter. Of course, in reality the orbital spatial extent is
of the same order as~$d$, so this multipole expansion is good
only for an order-of-magnitude estimate.
The overall charge neutrality and the vanishing dipole moment
of the unperturbed charge distribution around each metal atom
(due to its $D_{3h}$ symmetry) make the first non-vanishing
multipolar moment to be the quadrupole. The quadrupole field
decays with distance as $1/R^4$, so, even though each atom feels
the field from images of all other atoms, the dominant contribution
comes from the nearest two images which are due to the atom
itself. The electric field at the point $\vec{R}=0$ produced
by the two quadrupole images at $\vec{R}'=(0,0,\pm{d})$ is
given by
\begin{equation}\label{Ez0=}
\mathcal{E}_z(0)=-\frac{3}{2}\,
\frac{\zeta_1-\zeta_{-1}}{\vep{d}^4}\,Q_{zz},
\end{equation}
where $Q_{zz}$ is the quadrupole moment of the electronic cloud
around a single metal atom,
\begin{equation}
Q_{zz}\equiv-|e|\int\left(3z^2-R^2\right)\rho(\vec{R})\,d^3\vec{R}.
\end{equation}
Although the quadrupole moment of a metal atom deformed by the
crystal field differs from that of an isolated atom, we assume
them to be of the same order and estimate the latter.
The electronic configuration of tungsten is [Xe]$4f^{14}5d^46s^2$,
and four out of five outermost $d$ orbital states are filled,
according to the Hund's rule.
Since filled atomic shells have zero quadrupole moment,
$Q_{zz}$ for tungsten is determined by the single empty
$5d$~orbital which we take to be the $m=0$ one, as it is this
one that is known to give rise to the lowest conduction band
of tungsten dichalcogenides. The wave
function of this orbital can be approximated by a hydrogen-like
one with  $(n,l,m)=(5,2,0)$, which moves in the Coulomb potential
with effective charge $Z_{eff}=16.74$~\cite{Clementi1967}.
The average quadrupole moment of a hydrogenic $|n,l,m\rangle$
state~\cite{Bethe:book},
\begin{align}\label{Qzznlm=}
Q_{zz}=&-|e|\frac{a^2_B}{Z_\mathrm{eff}^2}\,
\frac{n^2[5n^2+1-3l(l+1)]}{2}\times \nonumber \\ &\times\frac{2l(l+1)-6m^2}{4l(l+1)-3},
\end{align}
then gives $Q_{zz}\approx{2}.75\,|e|a_B^2$ for tungsten
($a_B\approx{0}.53\:\mbox{\AA}$ is the Bohr radius).
For molybdenum, the electronic configuration in the gaseous
phase, [Kr]$4d^55s^1$, has a filled $4d$ shell with zero
quadrupole moment. However, given the fact that the lowest
conduction band of molybdenum dichalcogenides is formed
mostly by the $4d$ $m=0$ orbital, just like for tungsten
dichalcogenides, the electronic configuration is likely
to be changed by the crystal field to [Kr]$4d^45s^2$.
Then, repeating the same calculation with $(n,l,m)=(4,2,0)$,
$Z_\mathrm{eff}=12.44$~\cite{Clementi1967}, we obtain
$Q_{zz}\approx{1}.86\,|e|a_B^2$ for molybdenum.

If a TMDC monolayer suspended in vacuum is subjet to an
\emph{external} electric field $\mathcal{E}_z$, the field
inside the monolayer is $\mathcal{E}_z/\vep$. It is this
one that should be matched with $\mathcal{E}_z(0)$ from
Eq.~(\ref{Ez0=}) to obtain the effective external field
due to the substrate:
\begin{equation}
\mathcal{E}_z^\mathrm{eff}=
-\frac{3\vep(\vep_1+\vep_2)}{(\vep+\vep_1)(\vep+\vep_2)}\,
\frac{Q_{zz}}{d^4}.
\end{equation}
Taking the static dielectric constants
$\vep=7$ (see \cite{Berkelbach2013} and
references therein), $\vep_1=1$ (vacuum),
$\vep_2=4$ (silica), and $d=3.14\:\mbox{\AA}$,
we obtain $\mathcal{E}_z^\mathrm{eff}\sim-0.1\:\mbox{V/\AA}$
for tungsten dichalcogenides.
Taking a highly dielectric substrate, $\vep_2\gg\vep$,
one can increase the effective field by about a factor of~2.


\section{Electric field from the Maxwell equations and local field effects}
\label{app:Maxwell}

Here we solve the Maxwell equations in the presence of the
oscillating monolayer polarization
$\vec{P}e^{i\vec{q}\vec{r}-i\omega{t}}\delta(z)$
sandwiched between two semi-infinite media with dielectric
functions $\varepsilon_1$ and $\varepsilon_2$ occupying the
half-spaces with $z>0$ and $z<0$, respectively.
Due to the in-plane isotropy of the Maxwell equations
and of the susceptibility,
$\chi_{\alpha\beta}\propto\delta_{\alpha\beta}$,
$\chi_{\alpha{z}}\propto{q}_\alpha$,
we can assume
the wave vector $\vec{q}$ to be along the $x$ axis, without any
loss of generality. For the three-dimensional waves propagating
in the two media, the $z$ component of the wave vector is
given by $q_{1z,2z}=\sqrt{\vep_{1,2}\omega^2/c^2-q^2}$.

For the $s$~polarization, the fields in the incident, reflected and
transmitted waves are parametrized by their electric field amplitudes
$\mathcal{E}_\mathrm{i}$, $\mathcal{E}_\mathrm{r}$ and
$\mathcal{E}_\mathrm{t}$, respectively. The non-zero components of
the fields are given by
(we omit the common factor $e^{i qx-i\omega{t}}$)
\begin{align}
&\mathcal{E}_y=\left\{\begin{array}{ll}
\mathcal{E}_\mathrm{i}e^{-i q_{1z}z}+\mathcal{E}_\mathrm{r}e^{i q_{1z}z},&
z>0,\\
\mathcal{E}_\mathrm{t}e^{-i q_{2z}z},&z<0,
\end{array}\right.\\
&B_x=\left\{\begin{array}{ll}
\frac{cq_{1z}}{\omega}\left(\mathcal{E}_\mathrm{i}e^{-i q_{1z}z}
-\mathcal{E}_\mathrm{r}e^{i q_{1z}z}\right),&
z>0,\\
\frac{cq_{2z}}{\omega}\mathcal{E}_\mathrm{t}e^{-i q_{2z}z},&z<0,
\end{array}\right.\\
&B_z=\frac{cq}{\omega}\mathcal{E}_y.
\end{align}
The boundary conditions on the tangential components of the
electric and magnetic fields correspond to the continuity
of the electric field and a jump in the magnetic field due
to the surface current:
\begin{align}
&\mathcal{E}_\mathrm{i}+\mathcal{E}_\mathrm{r}
-\mathcal{E}_\mathrm{t}=0,\\
&\frac{cq_{1z}}\omega\,(\mathcal{E}_\mathrm{i}-\mathcal{E}_\mathrm{r})
-\frac{cq_{2z}}\omega\,\mathcal{E}_\mathrm{t}
=-\frac{4\pi i\omega}c\,P_y,
\end{align}
When the incident wave is absent, but the layer polarization
acts as a source producing the outgoing field, the amplitudes
of the latter are given by
\begin{equation}
\mathcal{E}_\mathrm{r}=\mathcal{E}_\mathrm{t}=
\frac{4\pi i\omega^2}{c^2(q_{1z}+q_{2z})}\,P_y.
\end{equation}

For the $p$ polarization, we choose the magnetic field amplitudes
$B_\mathrm{i}$, $B_\mathrm{r}$, $B_\mathrm{t}$, to parametrize the
fields, which are sought in the form
\begin{align}\label{TMa=}
B_y={}&{}\!\!\left\{\begin{array}{ll}
B_\mathrm{i}e^{-i q_{1z}z}+B_\mathrm{r}e^{i q_{1z}z},&z>0,\\
B_\mathrm{t}e^{-i q_{2z}z},&z<0,
\end{array}\right.\\
\label{TMb=}
\mathcal{E}_x={}&{}\!\!\left\{\begin{array}{ll}
\frac{cq_{1z}}{\varepsilon_1\omega}\left(-B_\mathrm{i}e^{-i q_{1z}z}
+B_\mathrm{r}e^{i q_{1z}z}\right),&z>0,\\
-\frac{cq_{2z}}{\varepsilon_2\omega}B_\mathrm{t}e^{-i q_{2z}z},&z<0,
\end{array}\right.\\
\label{TMc=}
\mathcal{E}_z={}&{}\!\!
\left\{\begin{array}{ll}
-\frac{cq}{\varepsilon_1\omega}\left(B_\mathrm{i}e^{-i q_{1z}z}
+B_\mathrm{r}e^{i q_{1z}z}\right),&z>0,\\
-\frac{cq}{\varepsilon_2\omega}B_\mathrm{t}e^{-i q_{2z}z},&z<0.
\end{array}\right.
\end{align}
The boundary conditions in this case are more subtle, because the
$z$-polarization $P_z\delta(z)$ produces an electrical double
layer. Let us assume $\delta(z)$ to be spread over a narrow
but finite region $|z|<d/2$ ($d$~being the monolayer thickness)
with a background dielectric constant~$\vep$.
In the first Maxwell equation for $\mathrm{div}\vec{D}$,
\begin{equation}\label{dzEz=}
i q\mathcal{E}_x+
\frac{\partial\mathcal{E}_z}{\partial{z}}=
-\frac{4\pi}\vep\,i qP_x\delta(z)
-\frac{4\pi}\vep\,\frac\partial{\partial{z}}\left[P_z\delta(z)\right],
\end{equation}
the first term is finite at $d\to{0}$. In the layer region it
can be neglected, which gives the following result for the
field in the layer region (we omit the terms which vanish at
$d\to{0}$):
\begin{equation}\label{Ez=delta}
\mathcal{E}_z(z)=-\frac{cq}{\vep\omega}\,B_\mathrm{t}
-\frac{4\pi}\vep\,i qP_x\!\!\!\int\limits_{-d/2}^z\!\delta(z')\,d z'
-\frac{4\pi}\vep\,P_z\delta(z),
\end{equation}
where the first term represents the field at $z=-d/2+0^+$,
related to the value at $z=-d/2-0^+$ from Eq.~(\ref{TMc=})
by the continuity of the normal component of the electric
displacement.

In the third Maxwell equation for $\rot\vec{B}$, the $\delta(z)$
term appears both in the polarization and displacement currents,
so it cancels out, and the only singularity in $B_y(z)$ comes from
the in-plane current,
\begin{equation}
B_y(z)=B_\mathrm{t}+\frac{4\pi i\omega}{c}\,P_x
\int\limits_{-d/2}^z\delta(z')\,d z'.
\end{equation}
However, from the Faraday's law,
\begin{equation}
\frac{\partial\mathcal{E}_x}{\partial{z}}-i q\mathcal{E}_z
=\frac{i\omega}{c}\,B_y,
\end{equation}
it follows that $\mathcal{E}_x$ must have a jump of
$-4\pi iqP_z/\vep$:
\begin{equation}\label{Ex=delta}
\mathcal{E}_x(z)=-\frac{cq_{2z}}{\varepsilon_2\omega}\,B_\mathrm{t}
-\frac{4\pi}\vep\,iqP_z\int\limits_{-d/2}^z\delta(z')\,d z'.
\end{equation}
This leads to the following boundary conditions for the amplitudes
in Eqs.~(\ref{TMa=})--(\ref{TMc=}):
\begin{align}
&\frac{cq_{1z}}{\varepsilon_1\omega}
\left(-B_\mathrm{i}+B_\mathrm{r}\right)
+\frac{cq_{2z}}{\varepsilon_2\omega}\,B_\mathrm{t}=
-\frac{4\pi iq}\vep\,P_z,\\
&B_\mathrm{i}+B_\mathrm{r}-B_\mathrm{t}=
\frac{4\pi i\omega}c\,P_x.
\end{align}
If we wish to determine the outgoing field produced by a source
layer polarization without the incident field, we have no
ambiguity; the amplitudes are given by
\begin{align}
&B_\mathrm{r}=\frac{4\pi i\omega}{c}\,
\frac{(q_{2z}/\varepsilon_2)P_x-(q/\vep)P_z}%
{q_{1z}/\varepsilon_1+q_{2z}/\varepsilon_2},\\
&B_\mathrm{t}=\frac{4\pi i\omega}{c}\,
\frac{-(q_{1z}/\varepsilon_1)P_x-(q/\vep)P_z}%
{q_{1z}/\varepsilon_1+q_{2z}/\varepsilon_2}.
\end{align}
However, we face a problem when we want to couple the field back
to the polarization, as we need the value $\mathcal{E}_{x,z}(z=0)$,
which is undetermined due to the singularity.
Here we note that coupling to the layer polarization is
in fact determined by $\int\delta(z)\,\mathcal{E}_{x,z}(z)\,dz$,
with the same $\delta(z)$ as in the spatial profile of the
polarization itself, which is nothing but the product of the
microscopic wave functions of the electron and the hole at
coinciding points
(see, e.~g., a microscopic treatment for excitons in a semiconductor
quantum well~\cite{Chen1988,Andreani1990}). In other words, we define
\begin{equation}
\bar{D}_{ij}(\vec{q},\omega)=\int\delta(z)\,D_{ij}(z,z';\vec{q},\omega)\,
\delta(z')\,d z\,d z',
\end{equation}
i.~e., as the projection on the polarization spatial profile in the
$z$~direction.
Then, the uncertainty due to the integral terms in Eqs.~(\ref{Ez=delta}),
(\ref{Ex=delta}) is resolved as
\begin{equation}
\int\limits_{-d/2}^{d/2}d z\int\limits_{-d/2}^zd z'\,
\delta(z)\,\delta(z')=\frac{1}2,
\end{equation}
by symmetry. On the contrary, the term
\begin{equation}
\int\limits_{-d/2}^{d/2}\delta^2(z)\,d z\equiv
\frac{\kappa_0}{2\vep}\sim\frac{1}d
\end{equation}
can only be determined from the microscopic theory.
The projected fields are given by
\begin{align}
\int\delta(z)\,\mathcal{E}_x(z)\,d z={}&{}
\frac{2(q_{1z}/\vep_1)(q_{2z}/\vep_2)}{q_{1z}/\vep_1+q_{2z}/\vep_1}\,
2\pi iP_x+{}\nonumber\\
{}&{}+\frac{q}\vep\,
\frac{q_{2z}/\vep_2-q_{1z}/\vep_1}{q_{1z}/\vep_1+q_{2z}/\vep_1}\,
2\pi iP_z,\\
\int\delta(z)\,\mathcal{E}_z(z)\,d z={}&{}
\frac{q}\vep\,
\frac{q_{1z}/\vep_1-q_{2z}/\vep_2}{q_{1z}/\vep_1+q_{2z}/\vep_1}\,
2\pi{i}P_x-\frac{2\pi\kappa_0}{\vep^2}P_z{}\nonumber\\
{}&{}+\frac{2(q/\vep)^2}{q_{1z}/\vep_1+q_{2z}/\vep_1}\,
2\pi iP_z.
\end{align}
To arrive at the final expressions (\ref{Dab=})--(\ref{Dzz=}),
one should recall that the field $\mathcal{E}_z$ which appears
in the definition of the dipole moment~(\ref{HRashba=}), and
which was used in the Rashba coupling constant estimate of
\cite{Kormanyos2014}, is not the local field, but the
external field, applied to a monolayer suspended in vacuum.
Hence, a factor $1/\vep$ should be absorbed into~$d_z$.

\section{Radiative rates in a parallel magnetic field}
\label{app:Zeeman}

As the $s$ and $p$ polarizations do not separate in a
parallel magnetic field, we return to the valley basis.
Namely, instead of using Eq.~(\ref{chipole=}), we
represent the susceptibility near each pole $E_\nu$ as
\begin{equation}\label{chipolevalley=}
\chi_{ij}(\vec{q},\omega)\approx
\sum_{\tau=\pm{1}}
\frac{\mu_{\tau{i}}^\nu(\mu_{\tau{j}}^\nu)^*}{E_\nu-\omega+i 0^+},
\end{equation}
where the vector $\boldsymbol{\mu}^\nu_\tau$, describes
the polarization of the exciton $\nu$ in the valley~$\tau$.
Then, the radiative self-energy can be approximated by
\begin{equation}\label{selfenergy=}
\Sigma_{\tau\tau'}^\nu(\vec{q})=-(\mu_{\tau{i}}^\nu)^*
\bar{D}_{ij}(\vec{q},E_\nu)\,\mu_{\tau'{j}}^\nu.
\end{equation}
For the dark $A$ excitons in the parallel magnetic field,
these vectors are given by
\begin{align}
&\boldsymbol{\mu}_{+1}\!=\!-i
\frac{ev\Phi(0)}{E_A^\mathrm{d}}\frac{g_\|\mu_BB_-}{4\Delta_\mathrm{c}}\left(\vec{e}_x+i\vec{e}_y\right)
+id_z\Phi(0)\,\vec{e}_z,\\
&\boldsymbol{\mu}_{-1}\!=\!i
\frac{ev\Phi(0)}{E_A^\mathrm{d}}\frac{g_\|\mu_BB_+}{4\Delta_\mathrm{c}}\left(\vec{e}_x-i\vec{e}_y\right)
+id_z\,\Phi(0)\,\vec{e}_z,
\end{align}
where $B_\pm=B_x\pm{i}B_y$. As we have seen in
Sec.~\ref{ssec:Gammaq}, the most important term that lifts
the valley splitting is the last term in Eq.~(\ref{Dzz=}).
Thus, we first consider the splitting at $\vec{q}=0$,
$\vec{B}_\|=0$, determined by the self-energy matrix
\begin{equation}
\Sigma^{(0)}(\vec{q}=0)=\frac{\Xi_0}{2}\left[\begin{array}{cc}
1 & 1 \\ 1 & 1 \end{array}\right],
\end{equation}
Its eigenvectors
determine the $z$-dipole-active state
$(1,1)^T/\sqrt{2}$ whose energy is shifted
up by $\Xi_0$, and the orthogonal one,
$(1,-1)^T/\sqrt{2}$, which is $z$-dipole-inactive.
In the case of Rashba-induced splitting at $\vec{q}\neq{0}$ these
states evolve into the $p$- and $s$-polarized ones, respectively.
At finite $\vec{q}$ and $\vec{B}_\|$, we project the self-energy
on the two eigenvectors $(1,\pm 1)^T/\sqrt{2}$,
which gives the radiative shift and the decay rate of the two states
\begin{align}
&\Omega_\perp(\vec{q})-\frac{i}2\,\Gamma_\perp(\vec{q})=
-8\pi{i}\,\Phi^2(0)\,\frac{e^2v^2}{E_A^2}
\left(\frac{g_\|\mu_BB_\|}{4\Delta_\mathrm{c}}\right)^2 \times \nonumber \\ &\qquad{}\times
\left[\frac{(E_A/c)^2}{q_{1z}^A+q_{2z}^A}\cos^2\tilde\phi
+\frac{(q_{1z}^A/\vep_1)(q_{2z}^A/\vep_2)}%
{q_{1z}^A/\vep_1+q_{2z}^A/\vep_2}\sin^2\tilde\phi\right]-{}\nonumber\\
&\quad{}{}-4\pi\Phi^2(0)\,\frac{evd_z}{E_A}\,
\frac{g_\|\mu_BB_\|}{4\Delta_\mathrm{c}}\,\frac{q_{1z}^A/\vep_1-q_{2z}^A/\vep_2}%
{q_{1z}^A/\vep_1+q_{2z}^A/\vep_2}\,q\sin\tilde\phi -{}\nonumber \\
&\quad{}-\frac{8\pi{i}q^2d_z^2\Phi^2(0)}{q_{1z}^A/\vep_1+q_{2z}^A/\vep_2}+\Xi_0,
\label{GammaBpar=}\\
&\Omega_\|(\vec{q})-\frac{i}2\,\Gamma_\|(\vec{q})=
-8\pi{i}\,\Phi^2(0)\,\frac{e^2v^2}{E_A^2}
\left(\frac{g_\|\mu_BB_\|}{4\Delta_\mathrm{c}}\right)^2\times \nonumber \\ &\qquad{}\times
\left[\frac{(E_A/c)^2}{q_{1z}^A+q_{2z}^A}\sin^2\tilde\phi
+\frac{(q_{1z}^A/\vep_1)(q_{2z}^A/\vep_2)}%
{q_{1z}^A/\vep_1+q_{2z}^A/\vep_2}\cos^2\tilde\phi\right],
\label{GammaBperp=}
\end{align}
where $\tilde\phi=\phi_\vec{q}-\phi_\vec{B}$,
and $\phi_\vec{q},\phi_\vec{B}$ are the polar angles of
$\vec{q}$ and $\vec{B}_\|$ in the $xy$ plane, respectively.

\section{Momentum integration of the decay rates}
\label{app:integral}

It is convenient to introduce a dimensionless variable
$s=c^2q^2/E_A^2$, then the integration over $\vec{q}$
transforms as $\int{d}^2\vec{q}=\pi(E_A/c)^2\int{d}s$,
and the radiative decay rates of the $A$~excitons,
given by Eqs.~(\ref{GammaTA=}), (\ref{GammaLAb=}),
(\ref{GammaLAd=}), (\ref{GammaBpar=}), (\ref{GammaBperp=}),
can be written as
\begin{align}
&\frac{\Gamma^\mathrm{b}_{TA}}{\Gamma_0}=
\Re\frac{2}{\sqrt{\vep_1-s}+\sqrt{\vep_2-s}},\\
&\frac{\Gamma^\mathrm{d}_{TA}}{\Gamma_0}=
\eta_\mathrm{R}\Re\frac{2s}{\sqrt{\vep_1-s}+\sqrt{\vep_2-s}},\\
&\frac{\Gamma^\mathrm{b}_{LA}}{\Gamma_0}=
\Re\frac{2\sqrt{\vep_1-s}\sqrt{\vep_2-s}}%
{\vep_2\sqrt{\vep_1-s}+\vep_1\sqrt{\vep_2-s}},\\
&\frac{\Gamma^\mathrm{d}_{LA}}{\Gamma_0}=
\eta_z\Re\frac{2\vep_1\vep_2s}%
{\vep_2\sqrt{\vep_1-s}+\vep_1\sqrt{\vep_2-s}},
\end{align}
where we keep only the dominant terms.
It is convenient to get rid of the square roots
in the denominators and take the real part of each
term in the numerators separately.
Then the integration becomes straightforward and gives
\begin{align}
&\int\frac{d^2\vec{q}}{(2\pi)^2}\,
\frac{\Gamma^\mathrm{b}_{TA}(\vec{q})}{\Gamma_0}=
\frac{(E_A/c)^2}{2\pi}\,\frac{2}3\,
\frac{\vep_2^{3/2}-\vep_1^{3/2}}{\vep_2-\vep_1},\\
&\int\frac{d^2\vec{q}}{(2\pi)^2}\,
\frac{\Gamma^\mathrm{d}_{TA}(\vec{q})}{\Gamma_0}=
\eta_\mathrm{R}\frac{(E_A/c)^2}{2\pi}
\frac{4}{15}\,\frac{\vep_2^{5/2}-\vep_1^{5/2}}{\vep_2-\vep_1},\\
&\int\frac{d^2\vec{q}}{(2\pi)^2}\,
\frac{\Gamma^\mathrm{b}_{LA}(\vec{q})}{\Gamma_0}=\frac{(E_A/c)^2}{2\pi}
\left[\frac{2}3\,\frac{\vep_2^{5/2}-\vep_1^{5/2}}{\vep_2^2-\vep_1^2}\right.-\nonumber \\
&\left.-\frac{2(\vep_1\vep_2)^{3/2}(\sqrt{\vep_2}-\sqrt{\vep_1})}%
{(\vep_2^2-\vep_1^2)(\vep_2+\vep_1)}-
\frac{2\vep_1^2\vep_2^2}{(\vep_2^2-\vep_1^2)(\vep_1+\vep_2)^{3/2}}
\right.\times\nonumber\\ &\times\left.
\ln\left(\sqrt{\frac{\vep_2}{\vep_1}}\,
\frac{\sqrt{\vep_2+\vep_1}+\sqrt{\vep_2}}
{\sqrt{\vep_2+\vep_1}+\sqrt{\vep_1}}\right)\right],
\label{FbTA=}\\
&\int\frac{d^2\vec{q}}{(2\pi)^2}\,
\frac{\Gamma^\mathrm{d}_{LA}(\vec{q})}{\Gamma_0}=
\eta_z\,\frac{(E_A/c)^2}{2\pi}
\left[\frac{2\vep_1^2\vep_2^2}{(\vep_1+\vep_2)^2}\right.\times \nonumber \\ &\times\left.
\frac{\sqrt{\vep_1\vep_2}+(\vep_1+\vep_2)/3}{\sqrt{\vep_1}+\sqrt{\vep_2}}+
\frac{2\vep_1^3\vep_2^3}{(\vep_2^2-\vep_1^2)(\vep_1+\vep_2)^{3/2}}
\right.\times\nonumber\\ &\times \left.
\ln\left(\sqrt{\frac{\vep_2}{\vep_1}}\,
\frac{\sqrt{\vep_2+\vep_1}+\sqrt{\vep_2}}
{\sqrt{\vep_2+\vep_1}+\sqrt{\vep_1}}\right)\right].
\label{FdTA=}
\end{align}
The function
$\mathcal{F}^\mathrm{d}_T(\vep_1,\vep_2)\equiv(4/15)
(\vep_2^{5/2}-\vep_1^{5/2})(\vep_2-\vep_1)$,
the function $\mathcal{F}^\mathrm{d}_L(\vep_1,\vep_2)$
is given by the square bracket in Eq.~(\ref{FdTA=}),
and $\mathcal{F}^\mathrm{b}(\vep_1,\vep_2)$ is given
by the half-sum of
$(2/3)(\vep_2^{3/2}-\vep_1^{3/2})(\vep_2-\vep_1)$ and
the square bracket in Eq.~(\ref{FbTA=}).


\begin{thebibliography}{99}
\bibitem{footnote_4}
Note that because $\vep_1\neq\vep_2$, for $q$ between
$\sqrt{\vep_1}E_A/c$ and $\sqrt{\vep_2}E_A/c$, one of
the square roots in $q_{1z}^A,q_{2z}^A$ is real, while the
other is imaginary, so separation of the real and imaginary
parts in (\ref{GammaTA=}), (\ref{GammaLAb=}), (\ref{GammaLAd=})
must be done with caution. We do not give explicit expressions
for $\Omega$ and $\Gamma$ separately, as they would be much
more bulky that the combined expressions for
$\Omega-(i/2)\Gamma$.
\bibitem{footnote_5}
In fact, lifting of the valley degeneracy for the dark $A$
excitons follows from a very simple and general consideration.
Consider the states at $\vec{q}=0$ in the absence of external
fields. Then, we have two degenerate excitations corresponding
to the $\pm\vec{K}$ valleys, whose transition dipole moments
are both directed along~$z$. Then, it is always possible to
form two
linear combinations of them, one of which will have zero dipole
moment and zero decay rate, while for the other one they will
be finite. If the decay rates for the two states are different,
so must be the energy shifts, which means lifting of the
degeneracy.
\end{thebibliography}
\end{document}